# Edge-Labelled Graphs and Property Graphs – a comparison from the user perspective

Paul Warren[a,b] and Paul Mulholland[a,c]

[a] Knowledge Media Institute, The Open University, Milton Keynes, MK7 6AA, U.K.
[b] paul.warren@open.ac.uk
[c] paul.mulholland@open.ac.uk

## Abstract

### Background

Property graphs and edge-labelled graphs are the two dominant paradigms for graph databases. The former are represented by proprietary languages, e.g. Cypher. The latter typically use the Resource Description Framework (RDF) and the SPARQL querying language, both standardized by the W3C. A proposal has been made to extend RDF to RDF* to enable an improved approach to reification, and to extend SPARQL to SPARQL* to enable querying of RDF*.

### Objective of study

This study compares participant acceptance of the two paradigms, as represented by Cypher and the proposed extensions to the W3C standards, i.e. RDF* and SPARQL*. Specifically, we compared preference for alternative modelling approaches, and accuracy of interpretation of queries.

### Method

The study used a web survey tool and attracted participants from across the world. The majority of the participants had some experience of graph database languages.

### Results

In general, modelling preferences are consistent across the two paradigms. When presented with location information, participants preferred to create nodes to represent cities, rather than use metadata; although the preference was less marked for Cypher. In Cypher, participants showed little difference in preference between representing dates or population size as nodes. In RDF*, this choice was not necessary since both could be represented as literals. However, there was a significant preference for using the date as metadata to describe a triple containing population size, rather than vice versa. There was no significant difference overall in accuracy of interpretation of queries in the two paradigms; although in one specific case, the use of a reverse arrow in Cypher was interpreted significantly more accurately than the ^ symbol in SPARQL. Based on our results and on the comments of participants, we make some recommendations for modellers.

### Future directions and future research in graph databases

Techniques for reifing RDF have attracted a great deal of research. Recently, a hybrid approach, employing some of the features of property graphs, has claimed to offer an improved technique for RDF reification. Query-time reasoning is also a requirement which has prompted a number of proposed extensions to SPARQL and which is only possible to a limited extent in the property graph paradigm. Another recent development, the hypergraph paradigm enables more powerful query-time reasoning. There is a need for more research into the user acceptance of these various more powerful approaches to modelling and querying. Such research should take account of complex modelling situations.

# 1   Introduction

Graph databases are a popular alternative to the relational database model. They particularly suit a range of applications which are inherently about networks, e.g. social networks, telecommunications networks, and logistics networks, see Robinson et al. [1], Chapter 5. They also avoid the pre-defined schemas employed by relational databases, and allow the schema to evolve [2].

There are two graph database formats: edge-labelled graphs and property graphs. The former has been standardised by the W3C as the Resource Description Framework (RDF) [3]. SPARQL [4], a query language for RDF, has also been standardised. As explained in more detail in Section 3.2, an extension of RDF, known as RDF*, has been proposed to enable reification; a parallel extension of SPARQL, known as SPARQL*, has been proposed to query RDF* [5]. There is currently no standardised format for property graphs, although an abstract model has been proposed [6]. However, one of the most common property graphs is Cypher [7], developed by neo4j[1]. Cypher is used in the study described in this paper. Based on a survey of open-source projects on GitHub, Seifer et al. [7] observed "higher activity in SPARQL related repositories". However, activity "in Cypher grew faster", i.e. Cypher, which appeared later than SPARQL, appeared to be catching up. In common with other property graph languages, Cypher is used both to create and to query graph databases.

The objective of the study described here is threefold:
- To compare the ease of use of the two language paradigms.
- To identify participants' preference for alternative semantically equivalent models; and, where appropriate, whether those preferences differed between the two languages.
- To determine where participants had particular difficulties in interpreting and querying models.

Our intention is to compare the conceptual differences between the two language formats, rather than being concerned with details of particular syntaxes. Therefore, with the exception of the question described in Section 11, the study is restricted to the chief features of the languages. Cypher has been chosen as an exemplar of property graph languages, rather than because of a specific interest in the details of its syntax.

In the remainder of the paper, Section 2 describes and compares the features of edge-labelled and property graphs, using the specific examples of RDF and Cypher. Section 3 describes the features of SPARQL and Cypher used in this study. Section 4 then describes how, under certain restrictions, it is possible to transform between the RDF* and property graphs. Section 5 describes related work. Section 6 elaborates on the motivation for this study and discusses the general methodology. Section 7 provides an overview of the study. Section 8 provides some information about the participants. The study was divided into five groups of questions, each group focussed on a particular modelling issue. Sections 9 to 13 describe these five groups, presenting an analysis of participant responses. Finally, Section 14 provides a concluding discussion and Section 15 makes some recommendations.

---

[1] https://neo4j.com/

# 2 Edge-labelled graphs and property graphs

This section provides an overview of edge-labelled and property graphs. The first two subsections describe the two formats. The third comments on terminology used in the two formats.

Angles et al. [8] provide a comprehensive review of graph query languages, both for edge-labelled graphs and property graphs. They include formal descriptions of the two formats. In this section, we draw on their descriptions, but for brevity restrict ourselves to an informal description.

We note here a key difference between the two approaches. Edge-labelled graphs, as implemented in the RDF standard, were designed to permit reasoning under an open world assumption. Property graphs, as exemplified by Cypher, were designed for querying at scale under a closed world assumption.

## 2.1 Edge-labelled graphs

Edge-labelled graphs consist of nodes and directed edges between pairs of nodes. All edges have a label, indicating the relationship between the two nodes connected by the edge. Edges can only have one label. However, there can be any number of edges between the same pair of nodes, each edge labelled differently. In general, the semantics of the edge label takes account of the directionality of the edge.

The RDF standard [3] states: "RDF graphs are sets of subject-predicate-object triples, where the elements may be IRIs, blank nodes, or datatyped literals". Subjects are the source and objects are the destination of the directed edge, which represents the predicate. Triples are terminated by a full stop (point). For this study, we ignore blank nodes. Subjects, predicates and objects can be IRIs. However, only objects can be literals. For the purposes of this study, IRIs are written as alphanumeric strings preceded by a colon. For simplicity, throughout this study, when writing RDF, we ignore the namespace identifier which would normally precede the colon. For the purposes of our study, literals are either integers, written in the normal way, e.g. 32; or character strings, written between single quotation marks, e.g. 'lawyer'. In the study, we use integers to represent years. For the benefit of those participants less familiar with RDF, we do not use the date datatype recommended in the standard [3].

## 2.2 Property Graphs

As there is no standard for property graphs, there are variations in definition and terminology between the various implementations. The following discussion is applicable to Cypher, and uses the terminology associated with Cypher.

As with edge-labelled graphs, property graphs consist of nodes connected by directed edges. All edges have a unique type, which is written preceded by a colon, e.g. *:worksFor* in Figure 2.1. Unlike with edge-labelled graphs, nodes and edges can have properties[2]; there is no limit to the number of properties which can be associated with a node or edge. For the purposes of this study, properties take a unique value, which will be either an integer, e.g. 2000, or a character string written between single quotation marks, e.g. 'Mary'[3]. When writing statements in Cypher, property names are separated from the associated value by a colon, and name and value are enclosed in braces. When a node or edge has several properties, then they are separated by commas and enclosed in the same set of braces. Properties are shown for both nodes and edges in Figure 2.1. The figure represents the information that a person with name 'Mary' is a lawyer and, from 2000 until 2010 worked for a company with name 'BuildCo', which is located in York. Note that property values are not unique identifiers. Referring to Figure 2.1, there may be several people called Mary, several companies called BuildCo, and several locations called York[4].

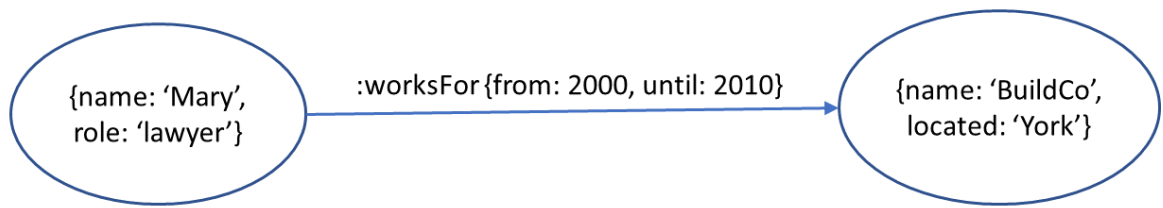

**Figure 2.1 Illustrating node and edge properties, and property type**

Nodes may have one or more labels, although the use of a label is not obligatory. In this study we restricted nodes to at most one label. Labels are preceded by a colon. Figure 2.2 illustrates a single node, with label *:Dog*. The label is commonly used to indicate membership of a class of objects. The use of multiple labels enables membership of multiple classes. However, there is no facility to enable a class hierarchy, as with RDFS.

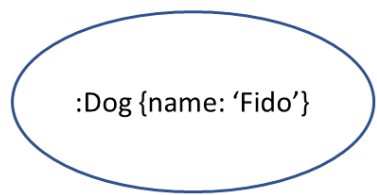

**Figure 2.2 Illustrating node label**

## 2.3 Terminology

It may be apparent from the previous two subsections that terminology differs between edge-labelled graphs and property graphs, and in particular between RDF and Cypher. The different uses of the word 'property' has already been noted. Another example is the use of *type* in Cypher to describe what would generally be called a *label* in an edge-labelled graph, and specifically a *predicate* in RDF; whereas in Cypher the term *label* is associated with nodes. In general, throughout this paper, when discussing RDF we use the terminology normally associated with RDF, and when discussing Cypher we use the terminology normally associated with Cypher.

---

[2] Note the difference in terminology between Cypher and SPARQL. A property in Cypher is quite different from a property in SPARQL, as used in the term 'property paths'. For this reason, we use the phrase 'predicate paths' to refer to what are normally referred to as property paths, e.g. in subsection 3.1.
[3] Cypher permits property values to take a number of additional primitive types which are not used in this study; see section 2.1 of [9].
[4] Cypher does associate a unique id to nodes and edges. These can be retrieved with the id() function. However, this is not recommended practice; in particular because the system reuses internal ids when a node or edge is deleted; see section 3.1.6 of [9].

## 3   Graph creation and querying

Subsection 3.1 lists the features of SPARQL used in this study; for brevity we do not describe these features in detail. Subsection 3.2 describes how RDF and SPARQL have been enhanced to create RDF* and SPARQL*. Subsection 3.3 describes the features of Cypher used in the study.

### 3.1   SPARQL – features used

With the exception of questions described in Section 11, our use of SPARQL is limited to SELECT and WHERE clauses, and predicate paths with the operators: concatenation (/), inverse (^) and one or more occurrences (+).

In Section 11 we make use of three functions for manipulating strings:
- *STR()* – this extracts the character string from an IRI; thus *STR(:London)* returns the string *':London'*. Note that the string returned includes the prefixed colon. Note also that when the argument to *STR()* is a variable, then the string returned is extracted from the IRI represented by the variable.
- *CONTAINS()* – this has two string arguments, and returns a logical true if the second argument is contained within the first, and a logical false otherwise. Thus *CONTAINS(':London', 'London')* will return a logical true. Note that one or both those arguments can be variables, where those variables represent strings.
- *FILTER()* takes one argument, which is either logical true or false. It has the effect of removing any solution for which the argument evaluates to false.

In subsection 11.2, these three functions are used together. Specifically, *FILTER (CONTAINS(STR(?group), ?groupname))* is used to restrict solutions to those for which the character string associated with the variable *?groupname* is contained within the character string forming the IRI associated with the variable *?group*.

### 3.2   RDF* and SPARQL*

At an early stage in the development of RDF, it was realised that there was a need to make statements about RDF statements, i.e. to create metadata describing triples, e.g. see Manola and Miller [10], Section 4.3. This process is known as *reification*. For a more recent comparison of various ways of achieving reification, see Hernández et al. [11]; in particular they compare query performance across four reification techniques and five SPARQL implementations. Hartig and Thompson [5] point out that, in general, reification techniques require a large number of triples and are cumbersome for querying. They propose a technique which mitigates these difficulties. Their approach, which they name RDF*, is an extension of RDF and backwards-compatible with RDF. Hartig [12] notes that RDF* is "simply a syntactic extension of RDF that makes dealing with statement-level metadata more intuitive".

In RDF*, triples can be enclosed within double angle brackets, and embedded within other triples. The following triple asserts that Adrian has worked for BuildCo from 2000:
    <<:Adrian :worksFor :BuildCo>> :from 2000 .

Triples can be embedded to an arbitrary depth. The following triple asserts that Adrian has worked for BuildCo from 2000, as an engineer:
    <<<<:Adrian :worksFor :BuildCo>> :from 2000>> :role 'engineer' .

Any embedded triple is regarded as being asserted. The previous triple can be regarded as the assertion of three triples, i.e. the outer triple plus the two inner triples:
    <<<<:Adrian :worksFor :BuildCo>> :from 2000>> :role 'engineer' .
    <<:Adrian :worksFor :BuildCo>> :from 2000 .
    :Adrian :worksFor :BuildCo .

A distinction can be made between metadata which is about the triple, and metadata which adds further information to the triple. An example of metadata about the triple, given in Hartig and Thompson [5], is the source of the triple, i.e. its provenance. Another example, given in Hartig [12], is the certainty of the triple, i.e. the probability of it being true.

Hernández et al. [11] give examples of the other use of reification, where further information is being added to the triple. Here they use the term *qualifier* to describe such metadata. Hartig [13][6] notes that RDF* can be used for both purposes. Our study is concerned with the second type of usage, i.e. the use of metadata as qualifiers.

SPARQL* extends SPARQL to enable the querying of RDF*. Just as RDF* is backwards-compatible with RDF, so SPARQL* is backwards-compatible with SPARQL. Just as RDF* embeds triples within triples using double angle brackets, so SPARQL* embeds triple patterns within triple patterns, using double angle brackets. Returning to the example earlier in this subsection, if we wished to know what role Adrian started with BuildCo in 2000, we could write:
    *SELECT ?role*
    *WHERE {<<<<:Adrian :worksFor :BuildCo>> :from 2000>> :role ?role}*

Because embedded triples are also asserted, we could also ask who Adrian works for:
    *SELECT ?company*
    *WHERE {:Adrian :worksFor ?company}*

Variables can appear within embedded triple patterns. For example, we can pose the very general query:
    *SELECT ?company ?startdate ?role*
    *WHERE {<<<<:Adrian :worksFor ?company>> :from ?date>> :role ?role}*

### 3.3  Cypher – features used

When writing Cypher, nodes are denoted by round brackets and edges by square brackets. Cypher graphs are created, and extended, using the *CREATE* clause and queried using the *MATCH* clause. The next two subsections describe these clauses. Subsection 3.3.3 then describes the WHERE keyword, which we use in a MATCH clause, and also testing for equality. Finally, subsection 3.3.4 describes two additional Cypher features which we use in one particular question.

---

[6] See comment of 17/08/2019.

### 3.3.1  *CREATE* clause

To create the graph of Figure 2.1, we could write:

*CREATE ({name: 'Mary', role: 'lawyer'})-[:worksFor {from: 2000, until: 2010}]-> ({name: 'BuildCo', located: 'York'})*

Here, neither of the nodes has a label, but the edge has an obligatory type (*:worksFor*). Both nodes and the edge have two properties, enclosed within braces. In the above statement, *name* is a property like any other, and does not denote a unique identifier. There can be other nodes with names *'Mary'* and *'BuildCo'* in the database. To illustrate the creation of a node with a label, the following statement creates the single node of Figure 2.2:

*CREATE (:Dog {name: 'Fido'})*

We can create multiple edges in one CREATE clause. For example, if we wish to create two edges, representing the facts that John is the brother of Mary and Stephen is married to Mary, we can write:

*CREATE ({name: 'John'}) –[:brotherOf]-> ({name: 'Mary'}) <- [:marriedTo]- ({name: 'Stephen'})*

This also illustrates that we can mix left-to-right and right-to-left edges in the same clause.

We can use a comma at the end of a line to join multiple lines into one *CREATE* clause. We might think, therefore, that we could achieve the same effect as above by writing:

*CREATE ({name: 'John'}) –[:brotherOf]-> ({name: 'Mary'}),*
      *({name: 'Stephen'}) –[:marriedTo]-> ({name: 'Mary'})*

However, because the value of the name property does not uniquely identify a node, this has the effect of creating two nodes with name *'Mary'*. If we want to indicate that there is one node with name *'Mary'*, then we need to use a variable:

*CREATE ({name: 'John'}) –[:brotherOf]-> (a {name: 'Mary'}),*
      *({name: 'Stephen'}) –[:marriedTo]-> (a)*

Here the use of the variable *a* indicates that the nodes at the end of each line are the same. Note that *a* is a variable with scope limited to the particular CREATE clause. It has no existence outside that clause, and it is not stored in the database.

### 3.3.2  *MATCH* clause

*MATCH* clauses use variables to indicate the node or edge of interest, and a *RETURN* keyword to indicate what values are to be output. In the example of the previous subsection, if we wish to know whom John is the brother of, we can write:

*MATCH ({name: 'John'}) –[:brotherOf]-> (a) RETURN a.name*

We can also associate a variable with an edge, and output edge property values. In the graph of Figure 2.1, if we wish to know the period during which Mary worked for BuildCo, we can write:

*MATCH ({name: 'Mary'}) –[a :worksFor]-> ({name: 'BuildCo'}) RETURN a.from, a.until*

Similarly to the use of the plus sign in SPARQL [14], Cypher uses an asterisk to indicate one or more occurrences of an edge. For example, assume we have an animal taxonomy represented by nodes with an associated property called group, and edges of type *:subGroupOf*. Then, if we wish to query for all the subgroups of *:Mammal*, and all the sub-subgroups, and sub-sub-subgroups etc., we could write:

*MATCH (a) –[:subGroupOf*]-> ({group: 'Mammal'}) RETURN a.group*

### 3.3.3   *WHERE* keyword and equals sign

We can use the *WHERE* keyword to impose additional constraints on a query. For example, suppose we have a graph containing the two nodes created by the following command:

*CREATE ({name: 'Joe', livesIn: 'London'}), ({name: 'Sue', livesIn: 'London'})*

We wish to output the name of the person who lives in the same city as 'Joe', i.e. 'Sue'. We can do this with:

*MATCH (a {name: 'Joe'}), (b) WHERE a.livesIn = b.livesIn RETURN b.name*

The equals sign is used to make a test for equality between two property values. We are assuming here that 'London' is a unique identifier. This is not guaranteed by Cypher; Sue and Joe may live in different towns with the name 'London'.

### 3.3.4   *labels()* function and *IN* keyword

The *labels* function takes as its argument a variable representing a node, and returns a list containing the string representations of the labels associated with the node, after removing the initial colon[7]. Because it returns a list, it can be used in conjunction with the *IN* keyword which tests for membership of a list. An example of this usage is given in subsection 11.2.

# 4   Mapping between edge-labelled graphs and property graphs

Hartig [12] formally presents mappings to transform between property graphs and edge-labelled graphs. For the former, he uses his own formalization of the property graph model. For the latter, he uses RDF*. In this section we describe these mappings informally. In subsection 4.1 we describe a mapping from RDF* to property graphs, and in subsection 4.2 a mapping in the reverse direction. Hartig [12] describes the RDF* to property graph transformation in subsection 4.1 as 'lossless', in the sense that the "resulting property graphs contains all information present in the original RDF* data". Hartig [12] also presents another

---

[7] Note the difference with the *STR()* function in SPARQL, which does not strip out the initial colon; see subsection 3.1.

RDF* to property graph transformation, which we present informally in subsection 4.3. This transformation is simpler and may appear more natural. However, the RDF* expressions which can be transformed are more limited, and also the transformation is not lossless. Whilst Hartig [12] describes the transformations with reference to an abstract model of property graphs, we use the Cypher notation in this section.

## 4.1 RDF* to RDF-like Property Graphs

Hartig [12] uses the phrase 'RDF-like Property Graphs' here to indicate that the resultant Property Graphs retain the structure of the RDF*. This is a requirement[8] for invertibility, i.e. that the original structure can be exactly recovered using the transformation described in the next subsection. The transformation represents each RDF triple as an edge in the Property Graph, with edge type determined by the RDF predicate. We use the IRI of the subject and object to provide values for a *name:* property associated with each node[9]. Consider, for example, the following RDF triple:

    *:Adrian :worksFor :BuildCo .*

The equivalent Cypher structure would be created with the following statement:

    *CREATE ({name: 'Adrian'})-[:worksFor]->({name: 'BuildCo'})*

This transformation can be extended to RDF*. Where a metadata triple has a literal as an object, then that literal becomes the value of an edge property. Consider an RDF* example from subsection 3.2:

    *<<:Adrian :worksFor :BuildCo>> :from 2000 .*

This would be translated into Cypher by creating two nodes, representing *:Adrian* and *:BuildCo*; and an edge of type *:worksFor* with a property *from:* with value *2000*. Thus, the Cypher statement to create this structure would be:

    *CREATE ({name: 'Adrian'})-[:worksFor {from: 2000}]->({name: 'BuildCo'})*

However, because RDF* is a more powerful representation than the property graph format, there are limitations on what can be transformed [12]:

    1. "The object of any metadata triple must be a literal."

This is because the object of a metadata triple is transformed into a property value, which must be a literal.

    2. "Metadata triples embed triples as their subject only (not as their object)."

The transformation algorithm breaks down in this case, since it requires that the object of the metadata triple be a literal.

---

[8] Although not a sufficient condition. See Hartig [12], Section 5.
[9] Hartig (2014) associates the IRI with an *:IRI* property. We use a *:name* property to align with the examples in our Cypher study.

3. "Metadata triples are not nested within one another."

Consider the example from subsection 3.2:

<<<<:Adrian :worksFor :BuildCo>> :from 2000>> :role 'engineer' .

Transformation to a property graph would require representing <<:Adrian :worksFor :BuildCo>> as a node in the property graph, which is not possible.

The impossibility of using Hartig's algorithm in certain situations does not mean that a sensible transformation is not possible. In the example given under condition 3 above, it would be possible for both *:from 2000* and *:role 'engineer'* to be property-value pairs for an edge with type *:worksFor*. We illustrate this later.

## 4.2 Property Graphs to RDF*

Hartig's [12] transformation from property graphs to RDF* transforms every edge to the predicate of an RDF triple, with start and end nodes transformed to subject and object of the triple. Any properties of the edge are transformed to metadata associated with the triple. Any properties of the nodes are represented by additional triples with the nodes as subject.

Taking a previous example:

*CREATE ({name: 'Adrian'})-[:worksFor {from: 2000}]->({name: 'BuildCo'})*

Following the example given in Hartig [12], subsection 3.3, this would become:

_:b1  :name 'Adrian' .
_:b2  :name 'BuildCo' .
<<_:b1 :worksFor _:b2>> :from 2000 .

Here *_:b1* and *_:b2* represent blank nodes. This does not recreate the original RDF* structure presented in subsection 4.1, which only had one metadata triple. To do that one would need to identify a particular node property, here *:name*, as providing the IRI of the node rather than being represented by an additional triple.

There is one condition which is required of the property graph to enable this transformation:
- There cannot be two edges of the same type, and with the same start and end nodes.

This requirement arises because the property graph model has no way of uniquely identifying two such edges[10]. This difficulty that this gives rise to is illustrated in subsection 13.1.

## 4.3 RDF* to Simple Property Graphs

Hartig [12] presents an alternative transformation from RDF* to property graphs which is simpler and more natural than the one presented in subsection 4.1. This transformation distinguishes between "*attribute triples*, that is, ordinary (non-metadata) triples whose object is a literal, and *relationship triples*, that is, ordinary triples whose object is an IRI or a blank node". Relationship triples are transformed in the same way as for the transformation of subsection 4.1. However, attribute triples are transformed to a single node, with a property-value pair representing the predicate and object. Thus the triple *:BuildCo :located 'York'* would be transformed to a node created with the following Cypher statement:

> *CREATE ({name: 'BuildCo', located: 'York'})*

---

[10] As noted in a footnote to subsection 2.2, Cypher does associate unique identifiers with each node and edge; however their use is discouraged.

Here we follow our convention of using the IRI of the subject to generate the value of the Cypher *name:* property. The IRI of the predicate is used to generate the value of the Cypher *located:* property.

This transformation requires one further restriction on the RDF*, in addition to those already discussed in subsection 4.1:
- Metadata triples must not contain embedded attribute triples.

This arises because, once the embedded triple has been transformed to a property graph node, there is no way of associating metadata with the properties of that node. To extend the previous example, consider the following metadata triple:

> *<<:BuildCo :located 'York'>> :since 2000*

Here there is no way of associating the *:since 2000* with the *located:* property in the property graph node. Had the transformation of subsection 4.1 been used, the inner triple would be transformed to a Cypher edge and the *:since 2000* would be transformed to an edge property.

## 5   Related work – database usability

During the initial development of database query languages, there were a number of studies investigating their usability. Reisner [15] provides an overview. She was concerned with methodology at a time when the application of behavioural science to computer science was a relatively new field; and in particular concerned to show that such an approach to usability studies was possible. Besides describing studies of query languages, e.g. comparing SQL with a number of other languages used at that time, she also describes studies comparing data models, i.e. hierarchical, relational and network. In one study [16] the relational system seemed to show a definite advantage; however, caution in interpreting this result was needed because of the effect of confounding differences in data model and differences in query language. This kind of methodological difficulty is frequent in studies of this kind; working with real languages and systems may give more immediately useful results, but at the same time results which are harder to interpret theoretically. She also describes a study [17] which avoided this problem by not using a query language, but rather comparing the comprehension of a relational and tree model. This study found that, at least for less experienced programmers, where there is a "natural tree structure", the tree model may be easier to comprehend.

Somewhat later, Catarci [18] describes how, starting as a database developer, she came to appreciate the importance of, and be involved in, usability studies. She comments that there were "very few empirical studies aimed at testing and validating the effectiveness of various query styles and interfaces". Based on her experience, she notes that requirements vary between users and between tasks. It is important, therefore, to ensure representativeness of users. Catarci was talking about involvement in the development process, but this applies also to usability testing. She also comments on the limited use of advanced functionality and a desire for simple interactions. Regarding the latter, she notes that some users were willing to give up functionality they had previously requested in the interests of simplicity of interaction.

More recently, Jagadish et al. [19] identified five "pain points" in database usability, caused by: difficulty in understanding joins between multiple tables; an excess of functionality and too many options; the appearance of, to the user, unexpected results; the need to issue a complete query before obtaining any feedback; and the difficulty of designing a schema or understanding a pre-existing one. They relate these difficulties to the difference between the user's mental model and the database model, and argue for a presentation data model closer to the natural model of the user's application. Depending upon the application this might be, for example, a geographic model, or it might be a network model. This is consistent with the use of graph databases in applications where the underlying user model is a network.

Batra [20] has looked at the factors which lead to complexity in data modelling. His analysis results in an eight dimensional framework for assessing how well tools can cope with complexity. In particular, he emphasizes complexity caused by the number of interactions, and their internal structure. Coping with this complexity requires heuristic shortcuts and, ideally, tools that provide mental aids. He notes the need to further investigate the relationship between structural and cognitive complexity. While chiefly concerned with the relational and entity-relationship models, much of his discussion is relevant to graph database modelling.

More recently still, Casterella and Vijayasarathy [21] were interested in the mistakes made by novices learning SQL. Although their primary interest was in comparing training strategies, their very detailed analysis of these errors has wider relevance. In particular, they distinguish between query structure errors and data model errors. The first arises through failure to correctly identify which clauses are needed in the query; the latter arises through failure to correctly identify which tables and columns are needed from the data model. More generally, they note that "novices struggle with analyzing natural language", i.e. translating from a query expressed in English to SQL. The kind of detailed analysis which Casterella and Vijayasarathy [21] have undertaken for SQL could also be applied to graph database languages. Indeed, to the extent that there are similarities between SQL and graph database languages, some of the observations of Casterella and Vijayasarathy [21] regarding the latter could be relevant to the former.

There have been a few studies looking specifically at graph database languages. Holzschuher and Peinl [22] have compared the readability of Cypher, Gremlin, SQL and a Java API. Based on a readability metric, they concluded that Cypher was the most readable of the four languages. Much of their paper is concerned with query performance, where Cypher also performed relatively well.

Warren and Mulholland [23] have investigated how accurately and rapidly users can reason about the main SPARQL features. Of relevance to this current study, questions with reverse predicates were answered less accurately and more slowly than analogous questions. The American philosopher C.S. Peirce, distinguished between symbolic signs, where representation depends on convention, and iconic signs, where representation depends on structural similarity [24][11]. In SPARQL, the use of ^ for reverse predicates is symbolic; whereas in Cypher the use of left-to-right and right-to-left arrows, to indicate the direction of the edge, is iconic. A relevant question is whether this leads to any significant difference in speed and accuracy of interpretation.

# 6 Motivation and methodology

The motivation for this study was to compare the edge-labelled and property graph paradigms, not from the technical, implementation viewpoint, but from the viewpoint of those creating and querying models. We were interested in how easily people could manipulate models and create queries in the two paradigms. Additionally, we were interested in how easily people could use RDF* and SPARQL*, i.e. the extensions to RDF and SPARQL proposed by Hartig and Thompson [5]. These two ambitions fit naturally together. Had we taken RDF as representing the edge-labelled paradigm, it would inevitably have suffered in expressive power compared with a property graph representation. However, the extension to RDF* brings the two paradigms closer to equality of expressiveness. Indeed, the discussion in Section 4 shows that RDF* can be regarded as the more expressive. This is not to say that we cannot represent in RDF any structure which can be created in a property graph model. The ability of the transformation of subsection 4.2 to map almost any property graph model to RDF*, combined with the fact that RDF* is purely a syntactic extension to RDF, shows that this is possible. However, without the syntactic extension offered by RDF*, such a mapping to RDF would require cumbersome reification techniques.

In this study, we use participants' rankings of models, and how accurately they identify correct and incorrect models and queries, as a proxy for their ability to think about and create models and queries. An alternative would have been to allow participants to create their own models and queries. A study of this sort would be a useful complement to our study. We opted for our format for two reasons. Firstly, were participants to create their own models and queries we would need to carefully assess participant responses, e.g. grading errors on a scale from trivial syntactic errors to major conceptual errors. Although possible, this approach would have been more difficult to implement. More importantly, we wanted to expose participants to some models and queries which they might not themselves have thought of. For example, with SPARQL* we wanted to understand the effect of reversing a query triple by placing the ^ operator in front of the predicate.

In Section 5 we noted the difficulty of confounding differences in modelling paradigm with differences in query language. The same danger is present in our study. Differences in the details of the RDF* / SPARQL* syntax and the Cypher syntax may be confounded with fundamental differences in the two paradigms. Our concern here is with the latter; which is not to say that the effect of differences of syntax are not in themselves worth studying. We chose to work with real languages so that all our models and queries could be tested for their correctness in real implementations. Specifically, we used Blazegraph[12] for RDF* / SPARQL* and Neo4j[13] for Cypher. As we were focussing on the differences in paradigms, we worked with subsets of SPARQL and Cypher. In addition, for RDF* and SPARQL* we avoided the use of namespace prefixes.

---
[11] There is also a third category, indexical signs, where representation depends on a physical connection. This category is not relevant to the current discussion.
[12] For a discussion of the RDF* / SPARQL* implementation in Blazegraph, see https://github.com/blazegraph/database/wiki/Reification_Done_Right
[13] https://neo4j.com/

If we had not chosen to use real languages, it would have been possible to avoid any differences in syntax which are not derived from the underlying paradigm differences. For example, SPARQL uses a SELECT statement, listing the variables to be output, at the beginning of a query, followed by a WHERE statement with the query pattern. A Cypher query starts with a MATCH statement with query pattern and ends with a RETURN keyword identifying the data to be output. Thus, not only does Cypher use different keywords from SPARQL, but it also reverses the order. It would have been possible, for example, to have designed a property graph language with the same keywords and the same order as SPARQL, but at the expense of not being able to empirically validate our models and queries.

# 7 Overview of study

This section is concerned with the organisation of the study. In subsection 7.1 we explain how the study was implemented. In subsection 7.2 we describe the overall structure of the study. In subsection 7.3 we describe the format used for the questions. Finally, in subsection 7.4 we describe how the results are analysed and presented. All the discussion in this section is relevant both to the RDF* / SPARQL* and Cypher versions of the study.

## 7.1 Organisation and implementation

We felt that it would be unrealistic to ask participants to answer questions on both paradigms. Hence there were two versions of the study, each with a different set of participants. Both versions were implemented using the Jisc online survey tool[14]. Participants were contacted through three routes. We used various specialist mailing lists to contact users of graph database languages. We also used internal mailing lists to contact computer scientists within our own organization. Finally, we directly emailed some appropriate personal contacts. Apart from providing a link to the study itself, we also provided participants with an electronic 'handout' which specified all they needed to know about the particular language being studied. Thus, there were two such handouts; one for RDF* / SPARQL* and one for Cypher[15]. The former contained eight pages; the latter six pages. In both cases the final page was a summary for quick reference. Participants were encouraged to look at the handout before starting the study and refer to it during the study.

For some questions we wanted to compensate for order effects by varying the order of questions in a section, or potential responses in a question, so that different participants saw different orders. For each of the two studies, we had four alternatives. Participants were given a link to a 'landing site' on a server maintained within our institution. Software on this landing site then redirected participants to one of the four alternatives on the Jisc server, cycling through each alternative in turn.

## 7.2 Structure of study

The study questionnaire began by providing some basic information, in part replicating some of the information in the handout. There was then a page to obtain the participants' agreement to take part in the study. The page began by noting that "This project has been reviewed by, and received a favourable opinion from, The Open University Human Research Ethics Committee, reference HREC/3568". There were then some details related to ethical issues, e.g. that anonymised data would be used to generate publications in the scientific literature. It was explained that clicking on 'Submit and Continue' to move on from this page would constitute agreement to the conditions of the study. During the study, as described later, participants had the opportunity to comment on the models and the study overall. Separately from overall permission to use the results of the study, we asked the participants whether they were happy for such comments to be quoted anonymously in research outputs. All participants responded that they were happy to be quoted in this way.

---

[14] https://www.onlinesurveys.ac.uk/
[15] The handouts are available at
https://ordo.open.ac.uk/articles/online_resource/Handout_for_RDF_and_SPARQL_study/12681737 and
https://ordo.open.ac.uk/articles/online_resource/Handout_for_property_graph_study/12681707.

The next part of the study asked the participants for some information about themselves. Firstly, basic personal information: age, gender, and in which country they worked. After this we asked about their expertise in four graph database languages: Cypher, Gremlin, GraphQL and SPARQL.

There were then two practice questions. These were to get participants used to the format of the questions. We stressed that their responses to these questions would not be analysed. The first question was concerned with data models; the second question with queries. For these questions, and for these questions only, after the participant had provided a response, potential answers were provided and the reasoning explained.

The next five parts of the study were concerned with the questions proper. These are described in Sections 9 to 13. In each case there was a question concerned with modelling. In one case, as described in Section 12, this was the only question. In the other four cases, there were then one or more questions concerned with queries.

Finally, we gave participants the opportunity to provide feedback about the study; asking what they found difficult, what they found easy, and giving them the opportunity to make any additional comments.

### 7.3 Format of questions

Subsection 7.2 explained that there were two types of question: modelling questions and querying questions. All modelling questions followed the same format, and all querying questions followed the same format. These two formats are illustrated in the next two subsections.

#### 7.3.1 Format of modelling questions

Figure 7.1 shows a screenshot of a modelling question, specifically for the question described in subsection 9.1. The screenshot shows the Cypher version; the RDF* / SPARQL* version follows the same pattern. At the top there is a description in English of the situation which it is required to model. There is then a query, in English, which we require our model to be able to answer. Below this are three models. In this case, the top model is correct and the other two are not; however in some cases there are more than one correct model. Then, at the left, the participant is required to indicate whether each of the models is correct or incorrect; for each model the participant must click on one, but only one of the options.

We also require that the participant ranks each model. Only one rank can be ascribed to each model. However, different models can be given the same rank. For example, it would be possible to rank all three models '1' or all three '3'. These two rankings may appear equivalent; we are saying that we think all three of equal quality. However, they do provide a different view on the models. A participant who ranks all three as '1' is likely to be implying not just that they are equal, but they are all good models. A participant who ranks all three as '3' is likely to be implying not just that they are equal, but that they are all bad models. In practice, in this example, a reasonable response is to rank model 1 as '1', since it is the only one which is correct, and rank the other two depending on one's views on their relative merits.

At the bottom right, participants are able to comment on each of the models. Whilst participants are required to specify the correctness and ranking of each model before being able to move on to the next page, there was no requirement to comment on each model.

As can be seen from Figure 7.1, participants were asked to "indicate which models are correct and capable of answering the query"; similar wording was used in the introductory page to the survey. However, several participants commented on a model, that whilst it was technically correct, they were marking it as incorrect because they thought it a bad model. It may well be that a number of other participants acted in the same way, without saying so explicitly. With hindsight, it is clear that the instructions should have been made more explicit so that participants were left in no doubt that, if the model accurately represented the situation and enabled the query to be posed, then it should be classified as correct, however bad it might be in the light of other criteria. In any case, the distinction between correctness and quality of model seems to have been blurred, at least for some participants, and this needs to be taken into account when interpreting the participants' responses.

### English

Adrian works as a lawyer for TransportCo. In addition, he works as an advisor for ArtsCo. Clare works as an accountant for TransportCo.

Required query:
- For which companies does Adrian work, and what role does he have in each company?

### Cypher models

```
(1) CREATE  (a {name: 'Adrian'})      –[:worksFor {role: 'lawyer'}]->      (b {name: 'TransportCo'}),
            (a)                        –[:worksFor {role: 'advisor'}]->     ({name: 'ArtsCo'}),
            ({name: 'Clare'})          –[:worksFor {role: 'accountant'}]->  (b)

(2) CREATE  (a {name: 'Adrian', role: 'lawyer'})       –[:worksFor]->  (b {name: 'TransportCo'}),
            (a {role: 'advisor'})                       –[:worksFor]->  ({name: 'ArtsCo'}),
            ({name: 'Clare', role: 'accountant'})       –[:worksFor]->  (b)

(3) CREATE  (a {name: 'Adrian'})       –[:worksFor]->  (b {role: 'lawyer', name: 'TransportCo'}),
            (a)                         –[:worksFor]->  ({role: 'advisor', name: 'ArtsCo'}),
            ({name: 'Clare'})           –[:worksFor]->  (b {role: 'accountant'})
```

This part of the survey uses a table of questions, view as separate questions instead?
Please consider the three models above.

| | Please indicate which models are correct, and capable of answering the query. | | Please rank each model (1 = best; 3 = worst). | | | Do you have any comments on this model? (optional question) |
|---|---|---|---|---|---|---|
| | correct | incorrect | 1 | 2 | 3 | |
| Model (1) | ○ | ○ | ○ | ○ | ○ | |
| Model (2) | ○ | ○ | ○ | ○ | ○ | |
| Model (3) | ○ | ○ | ○ | ○ | ○ | |

Submit and continue >

**Figure 7.1 Screenshot of a modelling question**

### 7.3.2 Format of querying questions

Figure 7.2 shows a screenshot of a querying question, specifically for the question described in subsection 9.2. Again, this is for the Cypher version, but the RDF* / SPARQL* version is similar. At the top there is a Cypher model. This is one of the models shown in Figure 7.1; in fact, the only correct one. Below this there is a query in English, and then three Cypher queries. Participants were asked to indicate whether each query is correct or incorrect. In this case, the top query is incorrect, the other two correct. For the querying questions, there was always at least one correct query and sometimes more than one.

[Figure showing Cypher model, English query, Cypher queries, and a survey table for indicating correct/incorrect queries]

**Figure 7.2 Screenshot of a querying question**

## 7.4 Analysis and presentation of results

For the modelling and querying questions, we calculate the proportion of accurate responses. A response is accurate if the participant properly classifies the model or query as correct or incorrect. For each model or query we display the proportion of accurate responses, e.g. see the upper part of Figure 9.2.

We also calculate the participants' mean preference for each model, on a scale of 0 to 1. For the situation in Figure 9.2, where there are three models, we equate a mean ranking of 1 with 1, a mean ranking of 3 with 0, and a mean ranking of 2 with 0.5. In general, where there are m models, a participant can rank each model on a scale of 1 to m. If a model has a mean ranking of x, then the preference is calculated as (m – x)/(m-1). This enables an easy comparison between the preference and the proportion of accurate responses.

Figure 9.2 and similar subsequent figures also show the 95% confidence intervals for each statistic. We use *R Studio*[16] and the *R* language [25] for all statistical calculations presented in this paper. For the accuracy data, because we are dealing with a binomial distribution, we use an exact method for calculating the confidence intervals[17]. Where there are only two models, so that they are each ranked '1' or '2', we use the same exact method for the preference statistic. Where there are more than two models, we calculate an approximate bootstrap confidence interval for the preference statistic[18]. When discussing statistical significance, we use the conventional 95% level.

---
[16] https://rstudio.com/
[17] Specifically, we use the *binom.confint()* function in the 'binom' package [26], with the exact (Pearson-Klopper) method.

## 8 Participants

This section provides some information about the participants. Subsection 8.1 provides information about the demographics and expertise of the participants. Subsection 8.2 discusses the relationship between participant expertise and accuracy of responses. Finally, in subsection 8.3 we compare how the two sets of participants performed.

### 8.1 The participants

We were interested in how widely our participants were distributed geographically; participants were asked to state the country in which they worked. In all, the participants were distributed between 18 different countries. Figure 8.1 shows, for each of the studies, the regional spread of the participants.

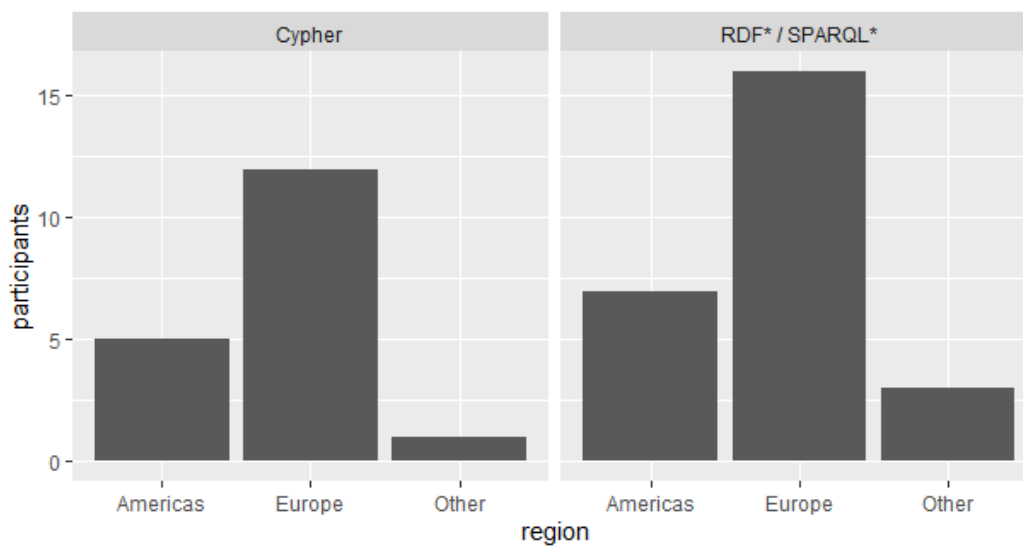

**Figure 8.1 Regional distribution of participants**

Participants were asked, optionally, to indicate their gender. For the RDF* / SPARQL* study there were 3 female participants, 17 male, and 6 who did not answer the question. For the Cypher study the corresponding numbers were 4, 12 and 2.

Participants were also asked to indicate their age range. Table 8.1 shows the age distribution of the participants for the two studies. Whilst we did not ask for the specific backgrounds of our participants, this data suggests that they included a high proportion of people experienced in their professions.

---

[18] Specifically, we use the *abc.ci()* function from the 'boot' package [27].

**Table 8.1 Distribution of participants by age in the two versions of the study**

|  | under 20 | 20 to 29 | 30 to 39 | 40 to 49 | 50 to 59 | 60 or over |
|---|---|---|---|---|---|---|
| **Cypher study** (N = 18) | 0% | 11.1% | 5.6% | 33.3% | 44.4% | 5.6% |
| **RDF*/SPARQL study** (N = 26) | 0% | 7.7% | 7.7% | 38.5% | 30.8% | 15.4% |

As explained in subsection 7.2, we asked people about their knowledge of four graph database languages: Cypher, Gremlin, GraphQL, and SPARQL. Using a four-point scale, Table 8.2 shows, for the participants in the Cypher study, their knowledge of Cypher and their knowledge of the four languages overall. For the latter, we took the maximum level of knowledge across the four languages for each participant. As can be seen, the knowledge overall was somewhat higher than for Cypher alone. This arose because three participants rated themselves as having more knowledge of SPARQL than of Cypher. Table 8.3 shows, for the participants in the RDF* / SPARQL* study, their knowledge of SPARQL and their knowledge of the four languages overall. Here we can see that there is no difference between the two distributions, i.e. no participant had any greater knowledge of the other three languages than of SPARQL.

**Table 8.2 Expertise of participants in Cypher study** (N = 18)

|  | No knowledge at all | A little knowledge | Some knowledge | Expert knowledge |
|---|---|---|---|---|
| **Cypher** | 38.9% | 11.1% | 33.3% | 16.7% |
| **Overall** | 33.3% | 5.6% | 38.9% | 22.2% |

**Table 8.3 Expertise of participants in RDF* / SPARQL* study** (N = 26)

|  | No knowledge at all | A little knowledge | Some knowledge | Expert knowledge |
|---|---|---|---|---|
| **SPARQL** | 0% | 11.5% | 19.2% | 69.2% |
| **Overall** | 0% | 11.5% | 19.2% | 69.2% |

Tables 8.2 and 8.3 show that the participants in the RDF* / SPARQL* study had, on average, considerably more knowledge of SPARQL than the participants in the Cypher study had of Cypher, or indeed any of the languages we asked about. To qualify this, the median for the distribution of Cypher knowledge in the Cypher study was between 'a little knowledge' and 'some knowledge'; the median for the distribution of SPARQL knowledge in the RDF* / SPARQL* study was 'expert knowledge'.

## 8.2  Participant expertise and response accuracy

We were interested in whether knowledge of the query language under study improved the accuracy of responses. We represent the overall accuracy for each participant as the total number of correct responses over all the questions. Table 8.4 shows the Spearman's rank correlation of accuracy against knowledge of Cypher, for the Cypher study. We show this overall, and for the modelling and querying questions separately. Similarly, Table 8.5 shows the Spearman's rank correlation of accuracy against knowledge of SPARQL for the RDF* / SPARQL* study.

**Table 8.4 Cypher study: accuracy versus knowledge of Cypher** (N = 18)

|  | Overall | Modelling questions | Querying questions |
|---|---|---|---|
| **Spearman's rank correlation** | 0.42 | 0.29 | 0.43 |

**Table 8.5 RDF* / SPARQL* study: accuracy versus knowledge of SPARQL** (N = 26)

|  | Overall | Modelling questions | Querying questions |
|---|---|---|---|
| **Spearman's rank correlation** | 0.04 | -0.14 | 0.31 |

In both studies, there was a positive correlation between the accuracy of answering the query questions and previous language knowledge. However, for the RDF* / SPARQL* study there was a negative correlation for the modelling questions, resulting in an overall correlation close to zero. In addition, as explained in more detail in the next subsection, the Cypher modelling questions were answered more accurately than the RDF* questions. These effects may arise because, as noted in subsection 7.3.1, some of the participants were marking a model as incorrect simply because they regarded it as a bad model. A comparison of Tables 8.2 and 8.3 shows that the RDF* / SPARQL* participants had greater expertise than the Cypher participants. It may be that this greater expertise led them to be more critical of the technically correct but badly designed models. It was also the case that all the RDF* models were correct, whereas some of the Cypher models were incorrect[19]. For these latter Cypher models, there is no ambiguity about how they should be marked. Additionally, Cypher is a language for creating and querying models; these models are created very much with querying in mind. RDF, on the other hand, is a framework for knowledge modelling for a range of purposes, e.g. communication and reasoning. This may lead to some RDF* models being viewed more critically.

## 8.3 Participant performance – a comparison between languages

We were interested to understand whether there is any overall difference in difficulty between the two paradigms. A comparison between the two studies is difficult for two reasons. Firstly, it is difficult to create two completely analogous studies. In particular, it is difficult to create comparable distractors, i.e. incorrect options. For one part of the study, described in Section 10, we were able to create strictly comparable questions. However, in general this was not possible. Secondly, it is clear from subsection 8.1 that the participants in the RDF* / SPARQL* study had appreciably more relevant knowledge than did the participants in the Cypher study. As explained below, we attempt to control for this.

Table 8.6 shows the mean accuracy for each study, overall and for the two sorts of questions. Overall, the mean accuracies are similar. However, when we look at the modelling and querying questions separately, we see that for the modelling questions the Cypher participants did better, whilst for the querying questions the RDF* / SPARQL* participants did better.

---

[19] In part this arose because RDF* permits more freedom in creating alternative correct models, in particular by permuting the nesting of the triples. In Cypher, where multiple properties can be associated with an edge, there are fewer correct models.

Possible reasons for the relatively poor performance on the RDF* modelling questions have been discussed in the previous subsection. A two-way Anova indicated a significant difference between question type ($F(1, 84) = 20.4372$, $p < 0.001$), no significant overall difference between languages ($F(1,84) = 0.3566$, $p = 0.552$), but a significant interaction effect ($F(1,84) = 4.2276$, $p = 0.043$)[20]. The interaction effect arises because the modelling questions were more accurately answered by the Cypher participants, whilst the querying questions were better answered by the RDF* / SPARQL* participants.

**Table 8.6 Accuracy of response in the two studies**

|  | **Overall** | **Modelling questions** | **Querying questions** |
|---|---|---|---|
| **Cypher study** | 0.88 | 0.83 | 0.90 |
| **RDF* / SPARQL* study** | 0.88 | 0.75 | 0.94 |

The ambiguous responses to the modelling questions mean that we cannot use these questions to compare the two paradigms. For the querying questions there is less ambiguity, and this makes it possible to use these questions to make a comparison. A difficulty is that the greater knowledge of the RDF* / SPARQL* participants may distort the comparison. We showed in subsection 8.2 that the correlation between accuracy and relevant knowledge was, for the querying questions, appreciable for both studies. Specifically, the Spearman's rank coefficient was 0.43 for Cypher and 0.31 for SPARQL*. Squaring these numbers gives the Coefficient of Determination, which represents the proportion of the variance in accuracy which can be explained by prior knowledge, i.e. 0.18 for Cypher and 0.10 for SPARQL*. It seems reasonable, therefore, to control for relevant knowledge, by using knowledge of Cypher in one case and of SPARQL in the other, as a factor in the analysis. This analysis showed a significant dependence on level of knowledge ($F(3, 37) = 3.6488$, $p = 0.021$), but no significant difference between the two paradigms ($F(1, 37) = 0.0399$, $p = 0.843$) and no interaction effect ($F(2, 37) = 0.2979$, $p = 0.744$)[21].

In summary, we have no evidence to suggest that, overall, there was any difference in comprehension between the Cypher and SPARQL query languages. We return to this topic for specific cases in subsections 9.2 and 10.2.

# 9 Metadata about predicates

This is the first of five sections which describe the survey questions and present an analysis of our results. The five sections are in the order in which the questions were presented to participants; broadly this was one of increasing model complexity. There were two questions in this part of the study; a modelling question and a querying question as described in the next two subsections. The underlying motivation for this section was to investigate how participants reacted to the use of metadata about predicates, as implemented in the two paradigms.

---

[20] Because there was a significant interaction effect, a type I Anova was used. However, changing the order of dependent variables makes no material difference to the conclusion.
[21] Because the data is unbalanced, i.e. there are different numbers of participants in each study, and because there is no significant interaction, we have used a type II ANOVA here.

## 9.1 Modelling

In this subsection we describe the question presented to each set of participants and then present and discuss the results.

### 9.1.1 Question

Figure 9.1 shows the modelling question. At the top is the model described in English, and the query we wish to answer, also in English. Below this are the three alternative models used in the Cypher study, and below this the three models used in the RDF* / SPARQL* study. For each model, a tick or a cross indicates whether the model is correct or incorrect. For both studies, model 1 associates the individuals' roles with the edge or predicate; model 2 associates roles with the subject, i.e. the individual; model 3 associates roles with the object, i.e. the company.

**English**

Adrian works as a lawyer for TransportCo. In addition, he works as an advisor for ArtsCo. Clare works as an accountant for TransportCo.

Required query:
- For which companies does Adrian work, and what role does he have in each company?

**Cypher models**

```
(1) CREATE (a {name: 'Adrian'})      –[:worksFor {role: 'lawyer'}]->   (b {name: 'TransportCo'}),
           (a)                       –[:worksFor {role: 'advisor'}]->  ({name: 'ArtsCo'}),
           ({name: 'Clare'})         –[:worksFor {role: 'accountant'}]-> (b)                        ✓

(2) CREATE (a {name: 'Adrian', role: 'lawyer'})   –[:worksFor]->   (b {name: 'TransportCo'}),
           (a {role: 'advisor'})                  –[:worksFor]->   ({name: 'ArtsCo'}),
           ({name: 'Clare', role: 'accountant'})  –[:worksFor]->   (b)                              ✗

(3) CREATE (a {name: 'Adrian'})   –[:worksFor]->   (b {role: 'lawyer', name: 'TransportCo'}),
           (a)                    –[:worksFor]->   ({role: 'advisor', name: 'ArtsCo'}),
           ({name: 'Clare'})      –[:worksFor]->   (b {role: 'accountant'})                         ✗
```

**RDF* models**

```
(1)   <<:Adrian :worksFor   :TransportCo>>   :role 'lawyer' .
      <<:Adrian :worksFor   :ArtsCo>>        :role 'advisor' .
      <<:Clare  :worksFor   :TransportCo>>   :role 'accountant' .                 ✓

(2)   <<:Adrian :role 'lawyer'>>       :worksFor :TransportCo .
      <<:Adrian :role 'advisor'>>      :worksFor :ArtsCo .
      <<:Clare  :role 'accountant'>>   :worksFor :TransportCo .                   ✓

(3)   :Adrian  :worksFor   <<:TransportCo :role 'lawyer'>> .
      :Adrian  :worksFor   <<:ArtsCo      :role 'advisor'>> .
      :Clare   :worksFor   <<:TransportCo :role 'accountant'>> .                  ✓
```

**Figure 9.1 Modelling question illustrating metadata about predicates**

In the Cypher question, only model 1 in which the metadata is associated with the edge, is correct. In fact, the Neo4j implementation rejects models 2 and 3 because it is not permitted to assign properties to a node in more than one occurrence of the node. In model 2, for example, we are not permitted to use the occurrence of the node with variable *a* to associate a *role* property. Although this was not made explicitly clear to the participants, it should be apparent that the association of the role with an individual breaks the connection with the company and makes it impossible to determine for which company Adrian worked as a lawyer and for which company he worked as an advisor. Similarly, the Cypher model 3 is incorrect because the role is associated with the company, and of the two people working for TransportCo, we cannot determine who works as lawyer and who as advisor. Hence, for both model 2 and model 3 we cannot answer the required query.

On the other hand, all the RDF* models are correct. Our supposition was that the most natural model is model 1, which associates the role information with the predicate. Individuals can have more than one role, as Adrian does in our example. Companies consist of people with many roles, as exemplified in our example by TransportCo, where Adrian is a lawyer and Clare an accountant. In model 2, although we associate the role with the individual in a triple, the position of this triple within a metadata triple makes clear the association between person, role and company. A similar argument applies to model 3. Using Hartig's transformation described in subsection 4.1, only model 1 is transformable to a property graph format. The RDF* model 2 is not transformable because the object of the metadata triple is not a literal, and model 3 is not transformable because the embedded triple is the object, rather than the subject, of the metadata triple.

In summary, in both studies the model which seemed to us most natural is model 1. In the Cypher study this was the only incorrect model. In the RDF* / SPARQL* study, whilst all three models were correct, only model 1 was transformable to Cypher using the transformations of Section 4.

### 9.1.2 Results and discussion

Figure 9.2 shows an analysis of the participants responses for both the Cypher and RDF* / SPARQL* studies. The models are labelled 1, 2 and 3 as in Figure 9.1. The top half of the figure shows the accuracy with which the models were identified as correct or incorrect. Considering as an example the top left quadrant, this shows that all participants accurately identified the first Cypher model as correct, whilst just under 70% and just over 80% accurately identified the second and third models as incorrect. The lines represent the 95% confidence interval for the estimated statistic. Turning to the top right quadrant, where all RDF* models were correct, it appears that an appreciable number of respondents failed to identify the second and third models as being correct. However, as we have already noted, some participants' comments suggest that models may be marked as incorrect because they were regarded as bad models, not because they were incapable of representing the situation and answering the required query.

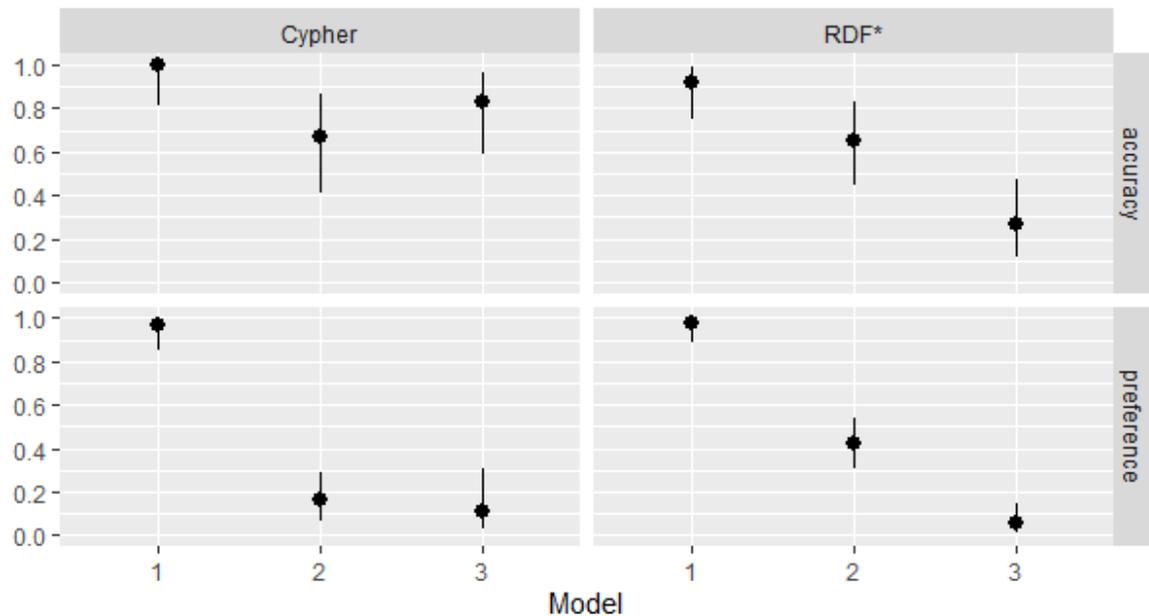

**Figure 9.2 Analysis of responses to modelling question shown in Figure 9.1**

The lower part of Figure 9.2 shows the mean preferences for the models, calculated from the rankings given to the models by participants; as with accuracy we use lines to represent the 95% confidence intervals. We display these preferences on a scale of 0 to 1, as discussed in subsection 7.4.

The figure shows that, for both studies, all or almost all participants regarded the first model as being correct and gave it highest preference. For the Cypher study, most participants gave the second and third models a low preference. Although the majority of participants understood that these two models were incorrect, there was an appreciable number who did not realise this. Amongst the RDF* models there was little liking for the third model, whilst the second model occupied second place. This is in line with the intuition that a role belongs to a person, not to a company.

Participants' comments regarding the Cypher models generally confirmed their awareness that models 2 and 3 were incorrect. A number of participants commented for these two models, that the role information was being overwritten, e.g. in model 2 the second line attempts to overwrite the role information for node *a* ('Adrian'). For the RDF* models, the responses confirmed the preference for model 1; one participant noting that "this one seems most intuitive". Regarding model 3, a participant commented "the model fulfils the query but the model is anyways not a good representation of the reality, therefore I say it is incorrect". We have already noted that comments of this sort need to be taken into account when interpreting the accuracy data for the models. Three participants were unhappy about the use of string literals for roles. A few commented on the relativeness closeness of each of the models to the original English. In response to model 3, one participant noted "if you were to change the name of :worksFor to :worksAs, this would match the English almost perfectly".

## 9.2  Querying

For the querying question, we took the one correct Cypher model from the previous question, along with its analogous RDF* model. We also used the same query as was presented in the modelling question.

### 9.2.1  Question

Figure 9.3 shows the query at the top, then the Cypher model and proposed Cypher queries, and finally the RDF* model and the proposed SPARQL* queries. In both cases, the first query is incorrect. In the Cypher case, the query incorrectly associates the role with the subject, i.e. Adrian. Similarly, for SPARQL* the role is associated with Adrian rather than the embedded triple. The other two queries are both correct. For both languages, the second query starts with Adrian, at the left of the query. The third query reverses the directionality. In the case of Cypher, this is achieved by reversing the *:worksFor* edge. In the case of SPARQL*, this is achieved by reversing the outer, *:role*, predicate [22].

**English query**

For which companies does Adrian work, and what role does he have in each company?

**Cypher model**

```
CREATE (a {name: 'Adrian'})   –[:worksFor {role: 'lawyer'}]->   (b {name: 'TransportCo'}),
       (a)                    –[:worksFor {role: 'advisor'}]->  ({name: 'ArtsCo'}),
       ({name: 'Clare'})      –[:worksFor {role: 'accountant'}]-> (b)
```

**Cypher queries**

| | | |
|---|---|---|
| (1) MATCH (m {name: 'Adrian'})–[:worksFor]–>(n) | RETURN n.name, m.role | ✗ |
| (2) MATCH ({name: 'Adrian'})–[e:worksFor]–>(n) | RETURN n.name, e.role | ✓ |
| (3) MATCH (n)<–[e:worksFor]–({name: 'Adrian'}) | RETURN n.name, e.role | ✓ |

**RDF* model**

```
<<:Adrian :worksFor   :TransportCo>>   :role 'lawyer' .
<<:Adrian :worksFor   :ArtsCo>>        :role 'advisor' .
<<:Clare  :worksFor   :TransportCo>>   :role 'accountant' .
```

**SPARQL* queries**

(1)
```
SELECT ?company ?role
WHERE {
    <<:Adrian :role ?role>> :worksFor ?company .
}
```
✗

(2)
```
SELECT ?company ?role
WHERE {
    <<:Adrian :worksFor ?company>> :role ?role
}
```
✓

(3)
```
SELECT ?company ?role
WHERE {
    ?role ^:role <<:Adrian :worksFor ?company>>
}
```
✓

**Figure 9.3 Querying question illustrating metadata about predicates**

---

[22] A more faithful analogue of the third Cypher query, would be to invert the embedded triple pattern, using the ^ operator, i.e. to write: *<<?company ^:worksFor :Adrian>> :role ?role*.  However, the Blazegraph implementation does not permit the use of ^ in embedded triple patterns.

### 9.2.2 Results and discussion

Figure 9.4 shows the accuracy with which the participants responded to the proposed queries, for the two studies. Most participants were able to identify the correct and incorrect queries. In fact, there is little difference between accuracy of response to all six of the proposed queries, with considerable overlap of the confidence intervals. The only appreciable difference is between the responses to query 3, i.e. the query where directionality was reversed. For the Cypher study, 94% of the participants realised this was a correct query, against 85% for SPARQL*. A one factor analysis of deviance does not indicate a significant difference between the two languages ($\chi^2(1) = 1.1076$, $p = 0.293$). However, it was noted in subsection 8.3 that the RDF* / SPARQL* study participants claimed more expertise in SPARQL than did the Cypher study participants in Cypher. Despite this, accuracy in the Cypher case was higher. A two-factor analysis of deviance, to control for language knowledge, revealed a significant difference between accuracy in the two studies ($\chi^2(1) = 5.2024$, $p = 0.023$), no significant effect of language knowledge ($\chi^2(3) = 6.0340$, $p = 0.110$) and no interaction effect ($\chi^2(2) = 0.0000$, $p = 1.000$)[23]. When we allow for the effect of prior knowledge, for this particular situation, the reversed predicate appears easier to interpret in Cypher than in SPARQL*.

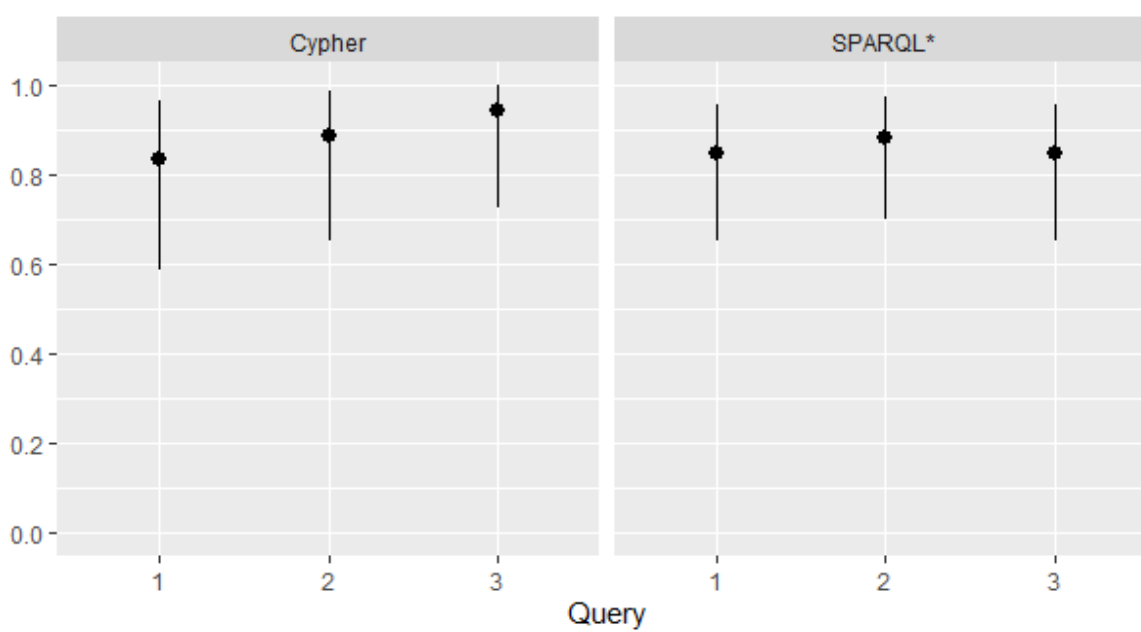

**Figure 9.4 Accuracy of responses to querying question shown in Figure 9.3**

## 10   Nodes, literals and reverse predicates

This section describes five questions. The first is a modelling question. The remaining four questions are querying questions which compare the arrow notation in Cypher with the ^ notation in SPARQL. The questions in this section do not involve the Hartig and Thompson [5] extensions to RDF and SPARQL.

---

[23] Similarly to subsection 8.3, we use a Type II analysis.

## 10.1 Modelling

This question compares two modelling approaches, based on a scenario composed of employees, their companies, and the companies' locations.

### 10.1.1 Question

Figure 10.1 shows the modelling question. For both the RDF* / SPARQL* and Cypher studies, there are two models. For the Cypher study, model 1 includes the location information as a property value, associated with the company node. For the RDF* / SPARQL* study, model 1 includes the location information as a literal, associated through a datatype property with a company. For model 2, the location is represented by a separate node or IRI. There are two required queries, and in each study, both models can correctly represent the information and answer the queries. The models are transformable using the transformation described by Hartig [12]. RDF* model 1 can be transformed to Cypher model 1 by using the transformation of subsection 4.3 for the triples with *:located* as a predicate, since these are attribute triples, i.e. their objects are literals. For the other three triples of model 1, and for all the triples of model 2, one would use the transformation of subsection 4.1.

For both studies, the order of the models was varied, so that some participants saw model 1 at the top, and some participants saw model 2 at the top.

**English**
Sophie works for CreativeCo, which is located in London. Brian works for BigCo, which is located in York. Diane works for AcmeCo, which is located in York.

**Required queries:**
- Where is the company located for which Brian works?
- Who works for a company located in the same town as Brian's company?

**Cypher models**

```
(1) CREATE ({name: 'Sophie'})    –[:worksFor]->    ({name: 'CreativeCo', located: 'London'}),
           ({name: 'Brian'})     –[:worksFor]->    ({name: 'BigCo', located: 'York'}),
           ({name: 'Diane'})     –[:worksFor]->    ({name: 'AcmeCo', located: 'York'})         ✓
```

```
(2) CREATE ({name: 'Sophie'})   –[:worksFor]-> ({name: 'CreativeCo'}) –[:located]->  ({name: 'London'}),
           ({name: 'Brian'})    –[:worksFor]-> ({name: 'BigCo'})      –[:located]-> (a {name: 'York'}),
           ({name: 'Diane'})    –[:worksFor]-> ({name: 'AcmeCo'})     –[:located]-> (a)         ✓
```

**RDF models**

```
(1)   :Sophie       :worksFor    :CreativeCo.
      :CreativeCo   :located     'London'.
      :Brian        :worksFor    :BigCo.
      :BigCo        :located     'York'.
      :Diane        :worksFor    :AcmeCo.
      :AcmeCo       :located     'York'.                    ✓
```

```
(2)   :Sophie       :worksFor    :CreativeCo.
      :CreativeCo   :located     :London.
      :Brian        :worksFor    :BigCo.
      :BigCo        :located     :York.
      :Diane        :worksFor    :AcmeCo.
      :AcmeCo       :located     :York.                     ✓
```

**Figure 10.1 Modelling question illustrating nodes and IRIs versus property-value pairs and literals**

### 10.1.2 Results and discussion

Figure 10.2 shows an analysis of the participants' responses. In each of the studies, the great majority of the participants identified that both models were correct. In each study there was a strong preference for model 2, in which the locations are represented by nodes or IRIs. A two factor analysis of deviance showed a significant difference in preference between the models ($\chi^2(1) = 55.741$, $p < 0.001$), no significant difference between RDF* and Cypher ($\chi^2(1) = 0.072$, $p = 0.788$), but a significant interaction effect ($\chi^2(1) = 5.418$, $p = 0.020$)[24]. This interaction effect indicates that the difference between the preferences for the two models was significantly less for Cypher than for RDF.

The participants' comments mostly confirmed the preference for model 2. A participant in the RDF* / SPARQL* study commented that model 1 "hinders future extensions of the database, as no attributes or relations can be added to strings". Another participant in this study commented "using strings to represent cities is such a bad modelling decision that this needs to be answered as an incorrect model". Similarly, in the Cypher study a participant commented that "locations are best treated as entities - and thus, nodes - in their own right". On the other hand, regarding RDF model 1, a participant did observe "in Property Graphs it is not uncommon to denote entities by simple strings".

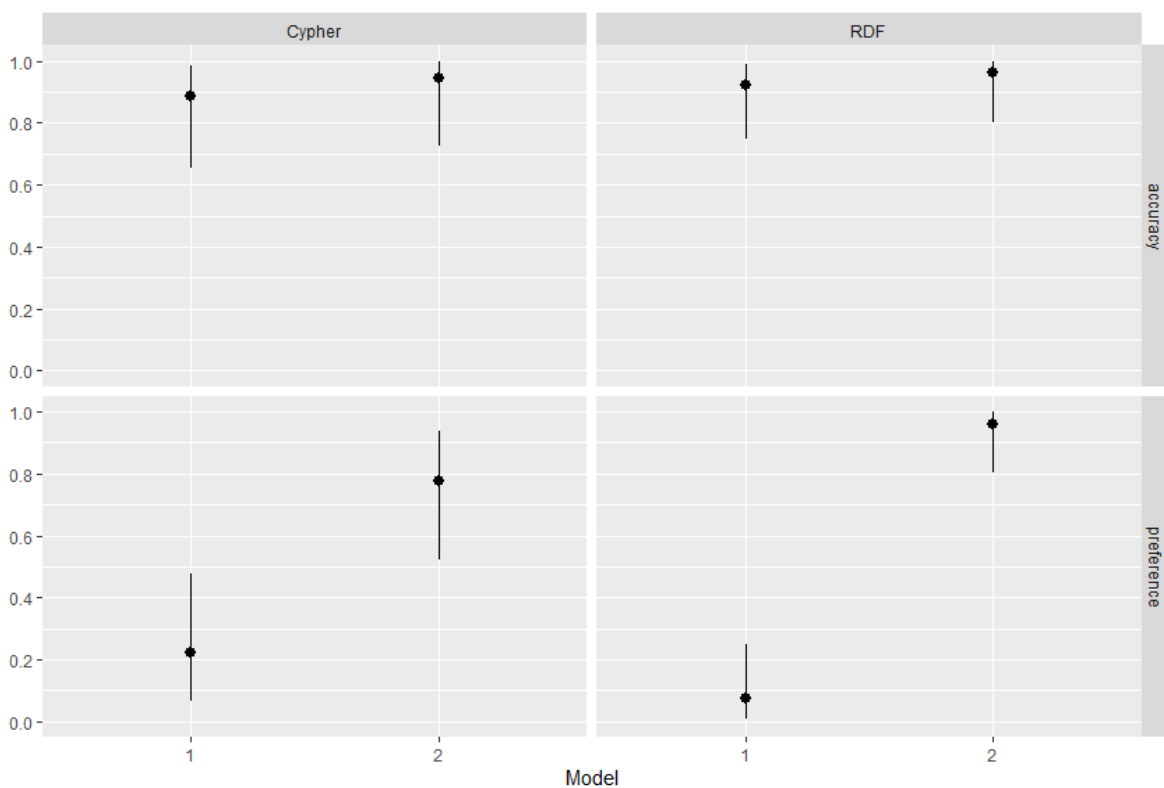

**Figure 10.2 Analysis of responses to modelling question shown in Figure 10.1**

---

[24] As there were two models, we can regard the distribution of rankings as following a binomial distribution, hence we use analysis of deviance. There being different numbers of participants in the two surveys, the analysis of deviance is unbalanced. We use type 1 analysis because of the high interaction effect. The p-values quoted were calculated with choice of model as the first factor. However, altering the order of factors makes no material difference to the conclusions.

## 10.2   Querying

There were four querying questions. For each of these questions, the Cypher and SPARQL forms were analogous. The object of these questions was to compare the iconic approach of Cypher, using forward and backward arrows to indicate the direction of the edges, with the symbolic approach of SPARQL which uses ^ to reverse the directionality of a predicate.

### 10.2.1   Questions

All four questions use model 2 from the previous modelling question. We have two levels of complexity, based on which of the two queries from the previous question we use. The first query ("Where is the company located for which Brian works?") we regard to be of lower complexity than the second query ("Who works for a company located in the same town as Brian's company?"), since the former requires fewer predicates or edges than the latter. At each level of complexity we have two questions, depending on whether the proposed queries start with a constant (Brian) or a variable. This pattern applies to both languages.

Figure 10.3 illustrates the questions. The figure is divided into three parts. At the top are the Cypher and RDF models. In the centre are the correct versions of the two lower complexity queries. For each question there were four potential queries presented to the participants; the correct one as shown in the figure, plus three incorrect. The four potential queries represented all the possible directions of the edge or predicate. Thus, for the incorrect Cypher questions of low complexity starting with a constant, we had:
- MATCH ({name: 'Brian'})–[:worksFor]–>()<–[:located]–(m) RETURN m.name
- MATCH ({name: 'Brian'})<–[:worksFor]–()–[:located]–>(m) RETURN m.name
- MATCH ({name: 'Brian'})<–[:worksFor]–()<–[:located]–(m) RETURN m.name

We took a similar approach for the Cypher question starting with a variable, and for the two analogous SPARQL questions.

The bottom part of Figure 10.3 shows the correct versions of the higher complexity queries, again in two forms: starting with a constant and starting with a variable. As with the lower complexity queries, for each of these two questions there were also three incorrect possibilities. These were generated by switching the directionality of the innermost two edges or predicates, the outermost two, and all four. For the SPARQL case, the higher complexity question starting with a variable had the following incorrect WHERE clauses:
- WHERE { ?person :worksFor / ^:located / :located / ^:worksFor  :Brian}
- WHERE { ?person ^:worksFor / :located / ^:located / :worksFor  :Brian}
- WHERE { ?person ^:worksFor / ^:located / :located / :worksFor  :Brian}

To compensate for order effects, for both studies the four questions were presented to participants in different orders, using the four 'landing sites' discussed in subsection 7.1.

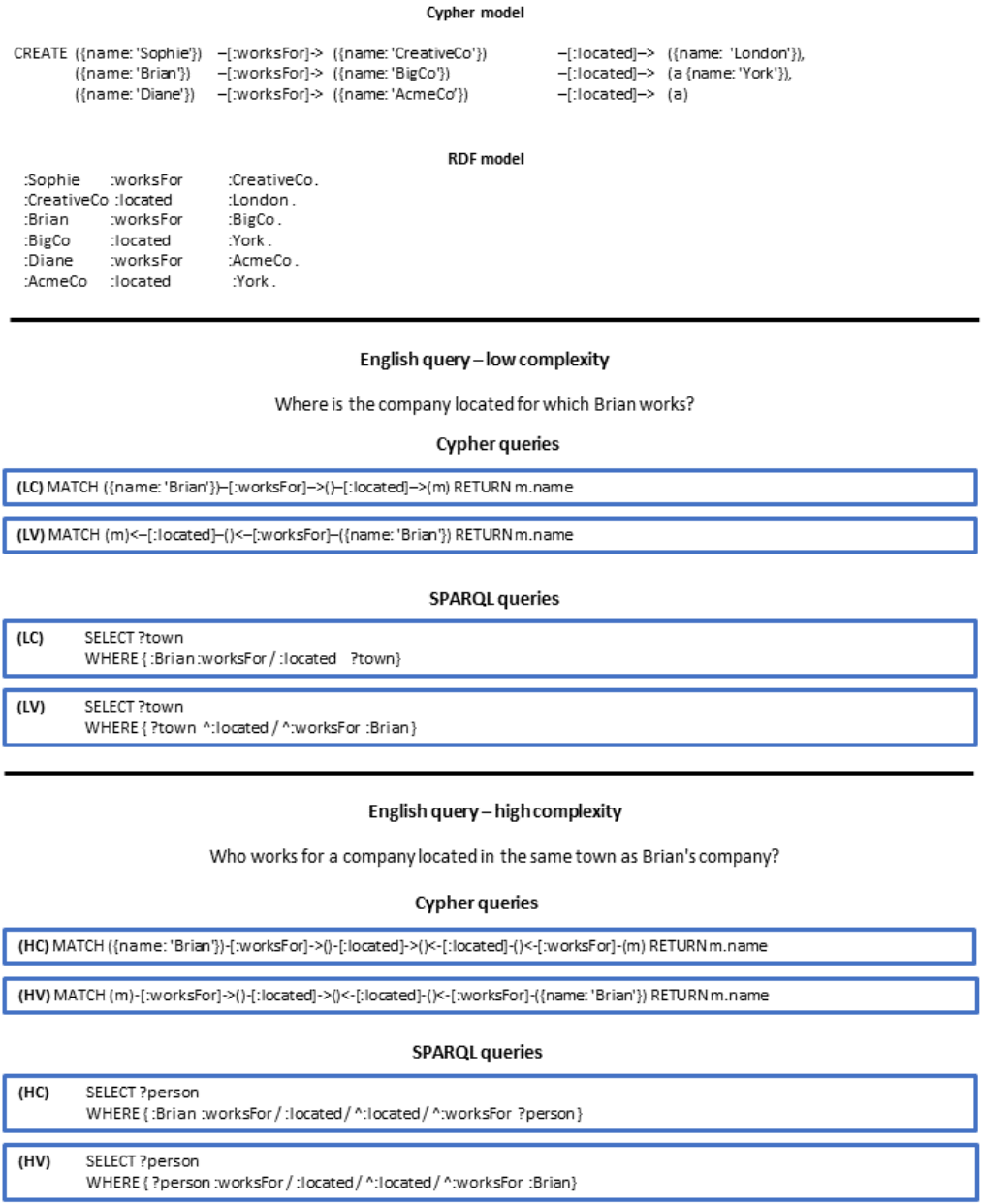

**Figure 10.3 Illustrating the correct responses for the four querying questions**
N.B. LC = low complexity, constant first; LV = low complexity, variable first
HC = high complexity, constant first; HV = high complexity, variable first

10.2.2 Results and discussion

Figure 10.4 shows an analysis of the participants' responses at the question level. For each question there are four potential responses. Hence for each question, each participant can be scored out of four. We label these questions as in Figure 10.3. As can be seen in Figure 10.4, the questions were answered with a high degree of accuracy. Moreover, there was little difference in accuracy between the questions. For each of the two studies, the total number of correct responses over the four questions, i.e. out of 16, was computed for each participant. A one factor ANOVA showed no significant difference between the two languages ($F(1, 42) = 1.0235$, $p = 0.318$). A two-factor ANOVA was also performed to control for the participants' expertise in the language being studied. This also showed no significant difference between the languages (F(1, 37) = 0.0095, p = 0.923). Nor was there a significant dependence on expertise (F(3, 37) = 0.9465, p = 0.428) or interaction effect (F(2, 37) = 0.3246, p = 0.725). In summary, for these questions there was no evidence of a significant different in accuracy between the iconic approach of Cypher and the symbolic approach of SPARQL.

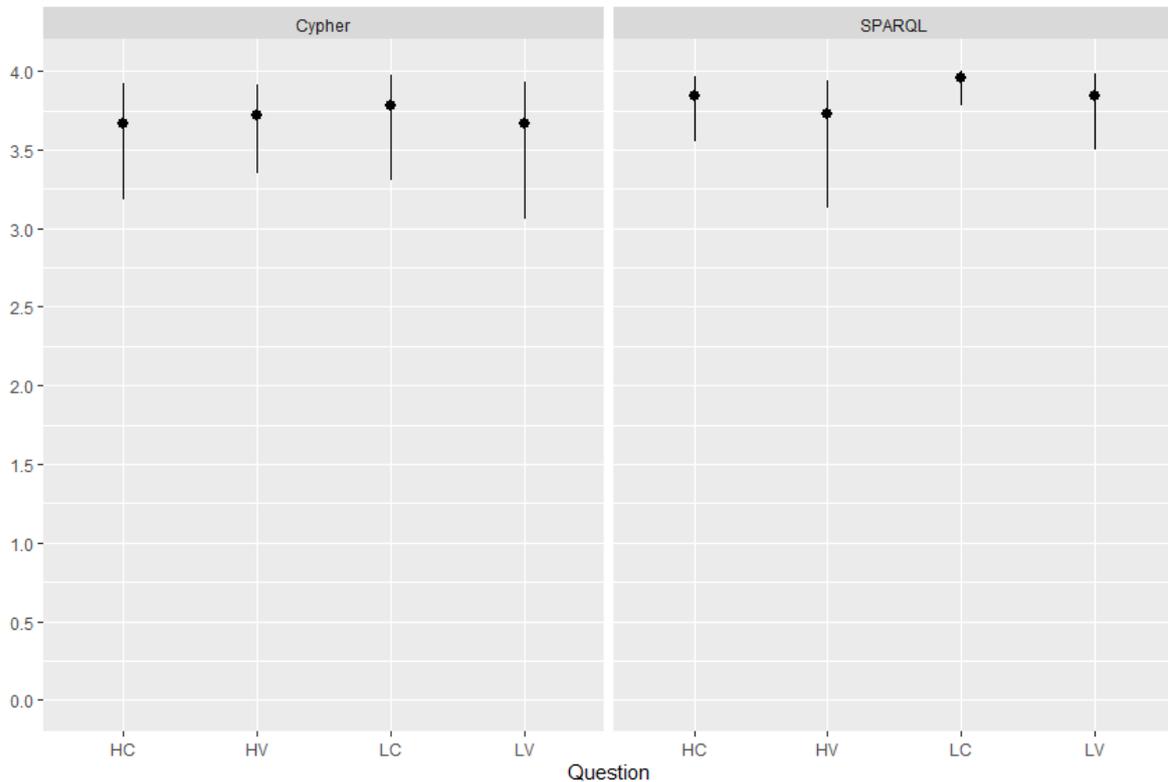

**Figure 10.4 Analysis of responses to querying questions show in Figure 10.3**
N.B. LC = low complexity, constant first; LV = low complexity, variable first
HC = high complexity, constant first; HV = high complexity, variable first

## 11     Class hierarchies

The questions in this section were concerned with modelling and querying class hierarchies. Like the questions in the previous section, they did not make use of the extended features of RDF* and SPARQL*. There was one modelling question, which asked participants to compare two models. These models were then used in two querying questions, with the same English query in both cases. To compensate for order effects, for both studies the two querying questions were presented to participants in different orders.

### 11.1    Modelling

The natural way to model hierarchies in RDF is using *rdfs:subClassOf*. Cypher has node labels, which are generally used to indicate class membership. Whilst a node can have an unlimited number of labels, indicating membership of many classes, there is no natural way of representing hierarchies. One way to create hierarchies in Cypher is to use the *labels()* function to extract the strings representing the labels of a node. Another way is to emulate the use of *rdfs:subClassOf* in RDF. For RDF we wanted to use the *rdfs:subClassOf* approach and we also wanted another approach which would be comparable to the use of the *labels()* function in Cypher. To achieve the latter approach, we created a model which mixed IRIs and strings. The corresponding query uses the *STR()* function to extract the string representation of an IRI.

### 11.1.1 Question

Figure 11.1 shows the models. For model 1, Cypher makes use of labels, e.g. *:Dog*, whilst RDF mixes strings and IRIs, so that both these models require string manipulations to correctly query them. For model 2, RDF takes the more natural modelling approach, whilst Cypher emulates this approach. We did not use *rdfs:subClassOf* but rather a predicate *:subGroupOf*, because participants might not necessarily be familiar with RDF, and we wanted to use a predicate name which fitted the context of the question[25].

The transformations of Section 4 do not permit the Cypher and RDF model 1 to be inter-converted, because the transformations do not take account of property graph labels. However, we can regard these models as being inter-convertible if we extend the transformations by transforming the Cypher label to the object of an RDF *:typeOf* predicate, and conversely transform the object of the RDF *:typeOf* predicate to a Cypher label. We also need to regard *:name* and *:group* as special kinds of Cypher properties, with values equated to RDF IRIs. The two versions of model 2 can be transformed using the transformations of Section 4.

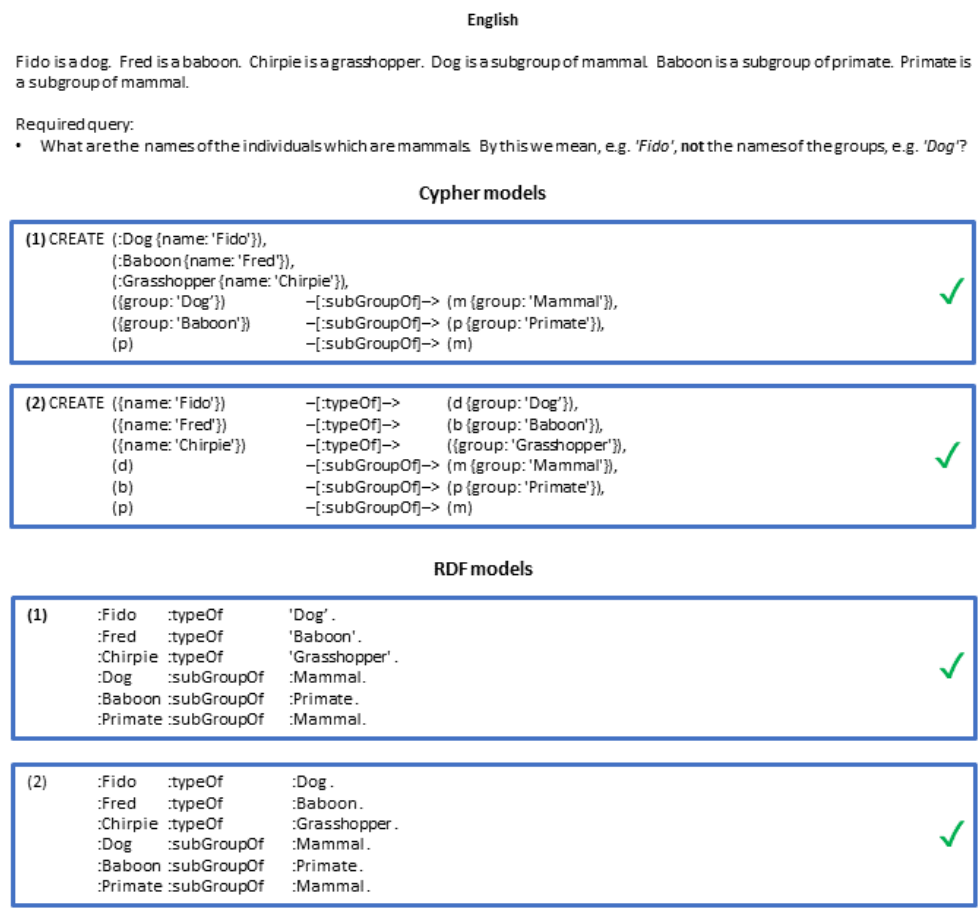

**Figure 11.1 Modelling question illustrating modelling of class hierarchies**

---

[25] In retrospect, this may have been a mistake. As already noted, all the participants in the RDF* / SPARQL* study were familiar with SPARQL, and hence with RDF, and the avoidance of *rdfs:subClassOf* may have caused confusion. There were, in fact, two comments along these lines, e.g. "The RDF vocabulary needs to be used".

For the RDF* / SPARQL* study, the order of the models was varied, so that some participants saw model 1 at the top, and some participants saw model 2 at the top[26].

### 11.1.2 Results and discussion

Figure 11.2 shows an analysis of the participants' responses. For the Cypher models, the majority of participants judged both models to be correct. For the RDF models, this was also true of model 2, whilst very few participants regarded model 1 as being correct. The contrast between model 1 in the two languages is interesting. Both require the construction of queries with textual manipulation, as is illustrated in section 11.2. In both cases, the elements needed to construct such a query were explained in the handout. As before, judgement whether a model was correct or incorrect may not always have been based on a literal interpretation of the question, but rather on a judgement on the quality of the model. A number of participants in the RDF* / SPARQL* study criticised mixing "resources and literal values for the same concept". Two explicitly stated that they appreciated that the model was not incorrect, but they regarded it as over complicated, and therefore marked it as incorrect. This may have less an effect on the Cypher models because the use of labels to denote class membership is natural in Cypher. One Cypher respondent noted "Neo4j generally pushes the use of labels for "types", but the path query for recursive subgroups is going to be awkward, requiring an equality condition on a node label and a node's property value." The approach used in model 2 is the natural way to define a class hierarchy in RDF; the majority of Cypher study participants also seemed happy to accept this approach.

Turning to the ranking of models, the effect is even stronger. For both languages, the majority of participants preferred model 2. This was the case in Cypher, even though the use of labels is the natural way to indicate class membership in Cypher. A two factor analysis of deviance illustrates this. There was no significant top-level effect between the two languages ($\chi^2(1) = 0.288$, $p = 0.593$), but there was a significant difference between the models ($\chi^2(1) = 59.027$, $p < 0.001$). There was a significant interaction effect ($\chi^2(1) = 6.158$, $p = 0.013$).[27] This is consistent with model 2 being the preferred model for both languages, but that preference being stronger for RDF.

---

[26] As a result of experimental error, the Cypher data for this modelling question relates only to the version with model 1 at the top. For this reason, this data only contained 10 participant responses, not the full 18. This will tend to create longer confidence intervals than for the other Cypher questions reported in this paper.

[27] As there was a significant interaction effect, Type I analysis was used. These values are taken from an analysis in which the model was specified first. Repeating the analysis in the alternative order makes no material difference to the conclusion.

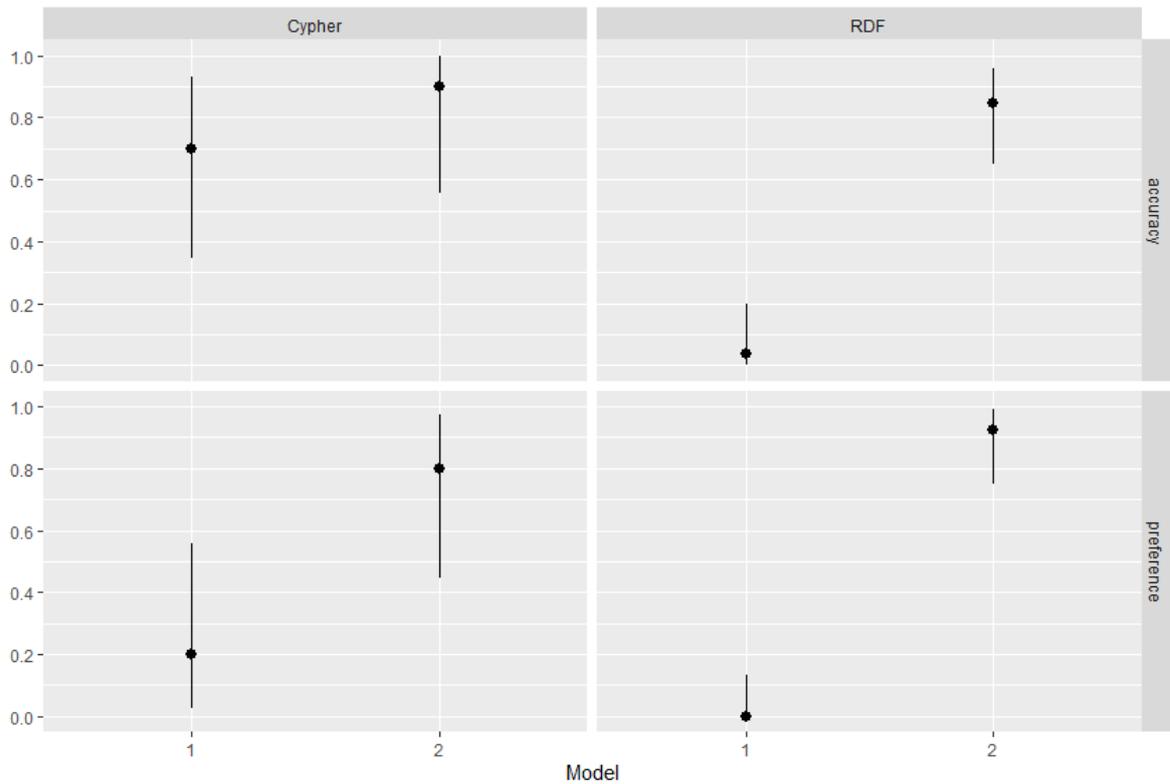

**Figure 11.2 Analysis of responses to modelling question shown in Figure 11.1**

## 11.2 Querying model 1

This question made use of model 1 in the previous subsection.

### 11.2.1 Question

Figure 11.3 illustrates the two variants of the question. For each language, query 1 is correct. The other two queries are incorrect, and in each case the two variants have been designed to be as similar as the two languages allow. In the case of query 2, the *:subGroupOf* predicate is used, but there is no recognition that, in order to make a comparison, we need to extract a string from, in the Cypher case a label and, in the SPARQL case an IRI. In the case of query 3, the need for string extraction is recognised, but there is no use of *:subGroupOf* to take account of the defined hierarchy.

**English query**

What are the names of the individuals which are mammals. By this we mean, e.g. *'Fido'*, **not** the names of the groups, e.g. *'Dog'*?

**Cypher model**

```
CREATE  (:Dog {name: 'Fido'}),
        (:Baboon {name: 'Fred'}),
        (:Grasshopper {name: 'Chirpie'}),
        ({group: 'Dog'})    –[:subGroupOf]–> (m {group: 'Mammal'}),
        ({group: 'Baboon'})–[:subGroupOf]–> (p {group: 'Primate'}),
        (p)                 –[:subGroupOf]–> (m)
```

**Cypher queries**

(1) MATCH (x), (y)–[:subGroupOf *]–>({group: 'Mammal'}) WHERE y.group IN labels(x) RETURN x.name   ✓

(2) MATCH (x)–[:subGroupOf *]–>({group: 'Mammal'}) RETURN x.name   ✗

(3) MATCH (x) WHERE 'Mammal' IN labels(x) RETURN x.name   ✗

**RDF model**

```
:Fido     :typeOf       'Dog'.
:Fred     :typeOf       'Baboon'.
:Chirpie  :typeOf       'Grasshopper'.
:Dog      :subGroupOf   :Mammal.
:Baboon   :subGroupOf   :Primate.
:Primate  :subGroupOf   :Mammal.
```

**SPARQL queries**

(1) SELECT ?name
    WHERE { ?name :typeOf ?groupname . ?group :subGroupOf+ :Mammal . FILTER (CONTAINS(STR(?group), ?groupname)) }   ✓

(2) SELECT ?name
    WHERE { ?name :subGroupOf+ :Mammal }   ✗

(3) SELECT ?name
    WHERE { ?name :typeOf+ ?groupname . FILTER (CONTAINS(STR(:Mammal), ?groupname)) }   ✗

**Figure 11.3 Querying question using model 1 from subsection 11.1**

### 11.2.2 Results and discussion

Figure 11.4 shows an analysis of the participants' responses. For Cypher, there were an appreciable number who failed to realise that query 1 is correct and query 2 is incorrect. Indeed, for both these queries, the number of accurate responses was 11 from the 18 participants, which was not significantly greater than chance ($p = 0.240$, exact one-sided test). Cypher query 3 and all the SPARQL queries were answered significantly better than chance; although there were an appreciable number of participants who failed to realise that SPARQL query 1 is correct. To compare the two languages, we calculated the number of accurate responses for each participant for each question, i.e. scoring each participant out of three. An ANOVA indicated a significant difference in performance between the two languages ($F(1,42) = 4.156$, $p = 0.048$). However, when we perform a two factor Type II ANOVA to control for prior knowledge, then the difference between the languages is no longer significant ($F(1,37) = 2.4872$, $p = 0.123$), whilst prior knowledge has a significant effect ($F(3,37) = 3.8511$, $p = 0.017$) and there is no interaction effect ($F(2,37) = 1.2898$, $p = 0.287$). In summary, the superior performance on the SPARQL queries is likely to be a result of the greater experience of the SPARQL participants.

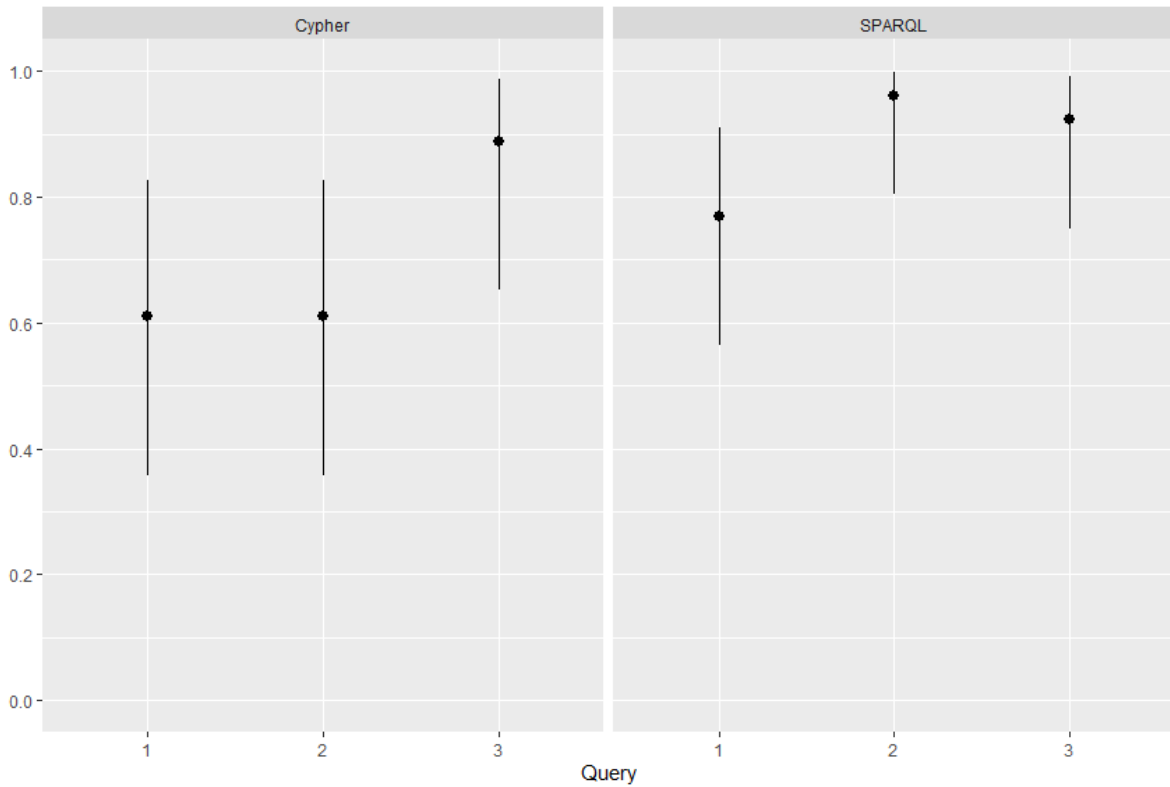

**Figure 11.4 Accuracy of responses to querying question shown in Figure 11.3**

## 11.3 Querying model 2

This question makes use of model 2 from subsection 11.1. The English query is the same as in subsection 11.2.

### 11.3.1 Question

Figure 11.5 illustrates the two variants of the question. For each variant, query 3 is the correct query, and the other two queries are incorrect. As with the previous question in subsection 11.2, for each incorrect query, the two variants have been designed to be as similar as the two languages allow. Query 1 uses the *:subGroupOf* predicate recursively, but without the *:typeOf* predicate to associate an individual animal with its lowest level group. Query 2, on the other hand, uses the *:typeOf* predicate, incorrectly using it recursively, but without the *:subGroupOf* predicate.

**English query**

What are the names of the individuals which are mammals. By this we mean, e.g. 'Fido', not the names of the groups, e.g. 'Dog'?

**Cypher model**

```
CREATE  ({name:'Fido'})      –[:typeOf]->      (d {group:'Dog'}),
        ({name:'Fred'})      –[:typeOf]->      (b {group:'Baboon'}),
        ({name:'Chirpie'})   –[:typeOf]->      ({group:'Grasshopper'}),
        (d)                  –[:subGroupOf]-> (m {group:'Mammal'}),
        (b)                  –[:subGroupOf]-> (p {group:'Primate'}),
        (p)                  –[:subGroupOf]-> (m)
```

**Cypher queries**

(1) MATCH (x)–[:subGroupOf *]->({group:'Mammal'}) RETURN x.name  ✗

(2) MATCH (x)–[:typeOf *]->({group:'Mammal'}) RETURN x.name  ✗

(3) MATCH (x) –[:typeOf]->()–[:subGroupOf *]->({group:'Mammal'}) RETURN x.name  ✓

**RDF model**

```
:Fido     :typeOf       :Dog.
:Fred     :typeOf       :Baboon.
:Chirpie  :typeOf       :Grasshopper.
:Dog      :subGroupOf   :Mammal.
:Baboon   :subGroupOf   :Primate.
:Primate  :subGroupOf   :Mammal.
```

**SPARQL queries**

(1) SELECT ?name
    WHERE { ?name :subGroupOf+ :Mammal }  ✗

(2) SELECT ?name
    WHERE { ?name :typeOf+ :Mammal }  ✗

(3) SELECT ?name
    WHERE { ?name :typeOf / :subGroupOf+ :Mammal }  ✓

**Figure 11.5 Querying question using model 2 from subsection 11.1**

### 11.3.2 Results and discussion

Figure 11.6 shows an analysis of the participants' responses. As with the previous question, we calculated the number of correct responses for each participant. On this basis, an ANOVA indicated no significant difference in performance between the two languages ($F(1,42) = 0.0018$, $p = 0.966$). This was confirmed when we performed a two factor Type II ANOVA to control for prior knowledge. In this case, neither the difference between the two languages ($F(1,37) = 0.2662$, $p = 0.609$), nor prior knowledge ($F(3,37) = 1.4438$, $p = 0.246$) was significant. In addition, there was no interaction effect ($F(2,37) = 0.2033$, $p = 0.817$).

However, this question was answered much more accurately than the previous question. For the previous question, the mean number of correct responses, calculated over both sets of participants, was 2.43. For this question, the corresponding mean number of correct responses was 2.89. A two-sided paired t-test indicated that this difference was significant ($t(43) = 3.3463$, $p = 0.002$). This is consistent with both sets of participants' preference for model 2, as described in subsection 11.1.

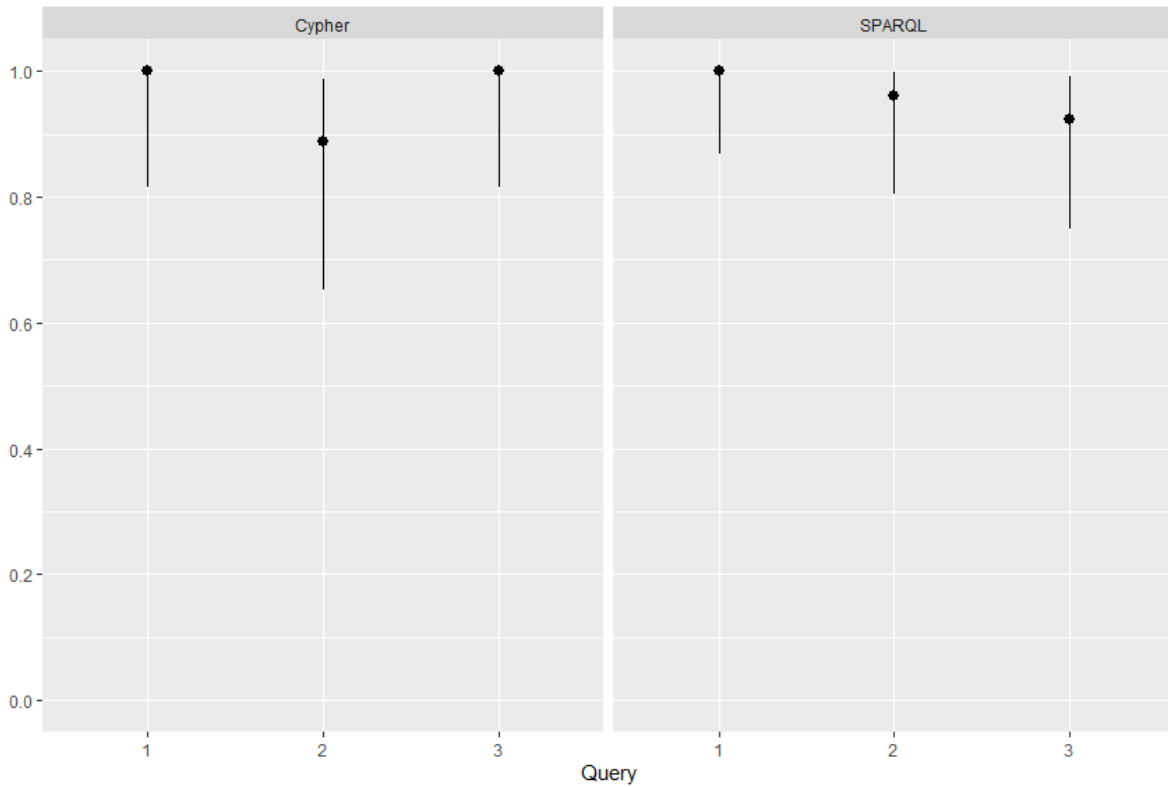

**Figure 11.6 Analysis of responses to querying question shown in Figure 11.5**

# 12    Metadata about node metadata

This section contained one question, a modelling question.

## 12.1    Modelling

The question was concerned with metadata about metadata, specifically attaching date-stamped population values to a node representing a city.

### 12.1.1  Question

Figure 12.1 shows the two variants of the question. This was the only question in the study where there were a different number of possible responses for the two languages. All the models are correct. For Cypher model 1, the size is associated with the edge, and the date with a node. For Cypher model 2, this is reversed. For Cypher model 3, both the size and date are associated with the edge. Cypher model 3 illustrates how an unlimited number of population-date pairs can be associated with a node, without creating any additional nodes. RDF* Models 1 and 2 use literals to represent the city population and the relevant date. We represented the dates by numeric literals, rather than date-specific datatypes, because we did not want to introduce the extra complication of such datatypes to participants who might have limited knowledge of RDF. Models 1 and 2 are both transformable between the two languages, using the transformations described in subsections 4.1 and 4.2. The RDF* models 3 and 4 are similar to model 2, except that the year is represented by an IRI rather than a literal. The only difference between models 3 and 4 is that, for the former, the IRI identifier contains the string 'year', to make clear what the IRI represents.

For both studies, the order of the models was varied, so that participants saw the models at different positions on the screen.

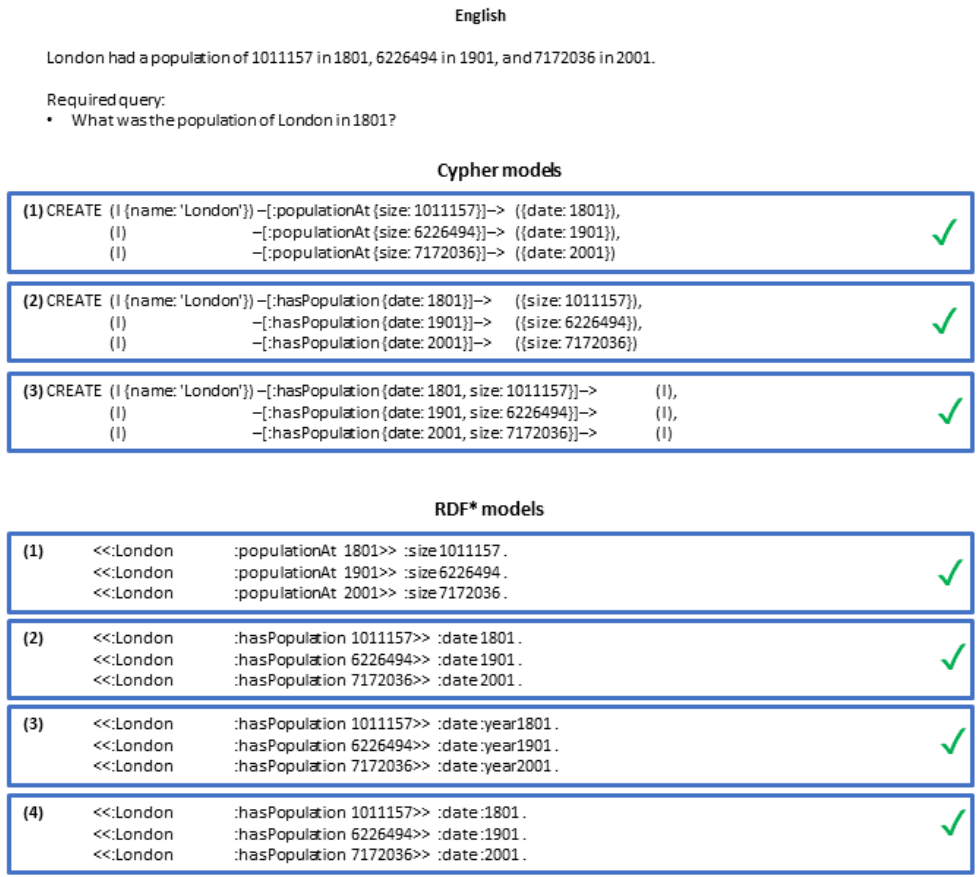

**Figure 12.1 Modelling question illustrating metadata about node metadata**

12.1.2 Results and discussion

Figure 12.2 shows an analysis of the participants' responses. For Cypher, there was little difference between the first two models, neither in accuracy nor in preference. One respondent, in criticizing Cypher model 1, commented: "It seems more intuitive that the value of population should be the size, not the year". On the other hand, another participant said of Cypher model 2: "But horrible... don't want a separate node for each number." There is a tension here. It seems more natural to make the object node the population, and to make the year a property. Yet, from a practical viewpoint, making the object node the year avoids arbitrarily creating nodes for particular numbers, and instead creates nodes for years which might be reused, e.g. for the population of other cities or events associated with the years. A resolution of this tension was suggested by one participant, who commented "Allowing for composite properties - the design of which is under way - would be the best option."

The third Cypher model was clearly less popular. The response on correctness of the model was 50%, i.e. exactly at chance. However, here again this may not reflect a literal answer to the question, but rather reflect on participants' opinion of the model. Indeed, one participant commented: "This can answer the query, but I'm marking incorrect as the *:hasPopulation* loop is intuitively not what the text represents". One participant looked more favourably on model 3, but made a suggestion which had some similarity to the comment above about composite properties: "The most correct and the best model, but really hard to read. I'd be arguing for nodes representing the measurement itself (e.g. node of type 'census report' or something similar, containing the date, size and information source.)"

A statistical analysis confirmed the intuitions from the figure. An ANOVA showed a significant difference between the rankings across the three models ($F(2,51) = 15.224$, $p < 0.001$). A subsequent Tukey honest-significance-difference (HSD) analysis shows no significance difference between models 1 and 2 ($p = 0.883$) but a significant difference between model 3 and models 1 ($p = 0.0001$) and 2 ($p = 0.00002$).

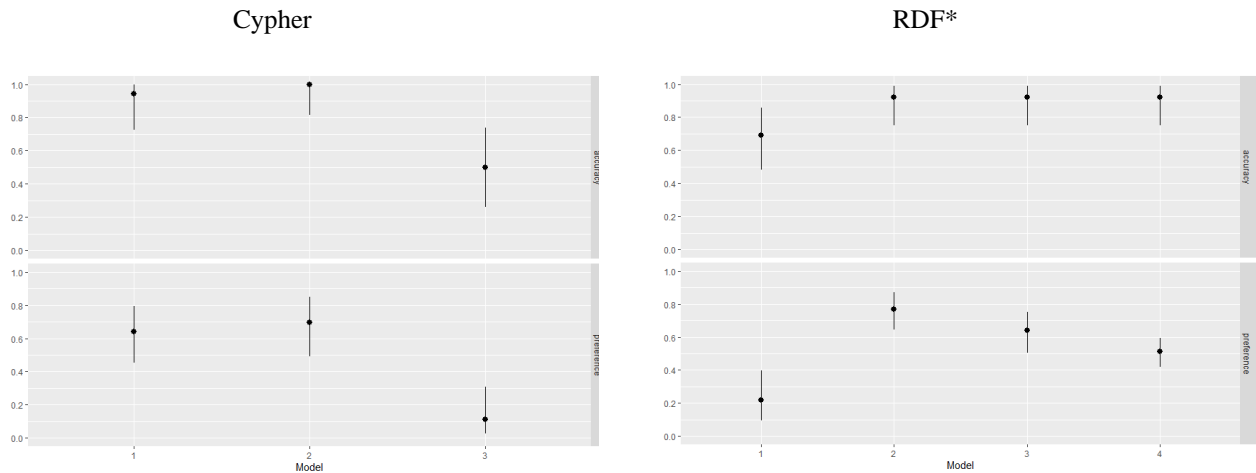

Figure 12.2 Analysis of responses to modelling question shown in Figure 12.1

For the RDF* models, the first model scored lower on accuracy and ranking than the other three. Once again, the lower accuracy score may reflect the dislike of the model shown in the ranking. The great majority of the participants recognised the other three models as being correct. The preference for models 2 to 4 over 1 is confirmed by a statistical analysis. An ANOVA of the rankings indicated a significant difference between the models ($F(3,100) = 14.527$, $p < 0.001$). A subsequent Tukey HSD analysis indicated a significant difference between model 1 and each of the other models (model 2: $p < 0.001$; model 3: $p < 0.001$; model 4: $p = 0.006$). Thus, there was a significant preference to have the date in the outer triple. Additionally, the Tukey HSD analysis showed a significant difference between models 2 and 4 ($p = 0.022$) but not between models 2 and 3 and models 3 and 4 ($p = 0.463$, in both cases).

Comments about the models varied. Consistent with the low ranking for model 1, one participant said of this model "A date is more logical to be a qualifying statement than the population. I would not use this." On the other hand, two other participants viewed it favourably, commenting: "this seems the natural order" and "This model allows for the simplest query." Two participants explicitly complained about the use of a literal for a date; another participant was unhappy that we did not use the RDF date datatype. On the other hand, one commented on model 4: "I think this should be just an integer". The particular choice of predicate names may influence responses. Commenting on model 1, one of the participants (who was also unhappy about the use of a literal) said: "but I would not mind experimenting with something like <<London :existsAt :year1801>> :hasPopulation XXXX".

The most striking result from this question is the significant preference amongst the RDF* models for the three which placed the date in the outer triple; compared with the lack of a significant preference amongst the Cypher models for the position of the date. For the RDF* models, it seems that the participants saw the date as "more logical to be a qualifying statement than the population". In general, this distinction did not seem to be a concern to the participants in the Cypher study.

## 13 Metadata about predicate metadata

This section contained two questions, a modelling question and a querying question. They made use of doubly-nested RDF* statements, and their analogues in Cypher.

### 13.1 Modelling

The question was concerned with an employee, his roles in a company, and the time period for which he held various roles. In the Cypher variant this allowed comparing the effect of placing the role and timing information as properties in the predicate and as properties in the object. In the RDF* variant, this allowed comparing the effect of placing the timing information in the middle and outer triple; and also comparing the positioning of the role and company information.

#### 13.1.1 Question

Figure 13.1 shows the modelling question. The first and fourth Cypher models cannot correctly answer the query. This arises because, in both cases, there is no way of associating, e.g. *from: 2000* with *role: 'engineer'* and *from: 2010* with *role: 'manager'* [28]. The other two models are correct, because the use of two edges enables the role and time information to be correctly associated. On the other hand, all the RDF* models are correct.

None of the RDF* models can be transformed to Cypher because of restriction 3 quoted in subsection 4.1:
- Metadata triples are not nested within one another.

Nor can Cypher model 2 be transformed to RDF*. This is because of the restriction given in subsection 4.2:
- There cannot be two edges of the same type, and with the same start and end nodes.

The example enables us to understand the reason for this restriction. Because we cannot uniquely identify an edge, it is not possible to maintain the association between, e.g. *role: 'engineer'* and *from: 2000*.

Cypher model 3 does not suffer from this problem, and in fact can be transformed to RDF*; although not to any of the RDF* models shown in the question. Such a transformation requires that the two object nodes, i.e. containing the role and time information, be transformed to blank nodes.

For both studies, the order of the models was varied, so that participants saw the models at different positions on the screen.

---

[28] In fact, in the Fuseki implementation, the first occurrence of a property value, e.g. *role: 'engineer'*, is overwritten by the second, i.e. *role: 'manager'*, so that only the latter appears in the model.

**English**

Adrian worked as an engineer for BuildCo from 2000 until 2010. Adrian has worked as a manager for BuildCo from 2010.

Required query:
- What were Adrian's roles at BuildCo, and when did he start them?

**Cypher models**

(1) CREATE ({name: 'Adrian'}) –[:worksFor {role: 'engineer', from: 2000, until: 2010, role: 'manager', from: 2010}]–>({name: 'BuildCo'}) ✗

(2) CREATE (a {name: 'Adrian'})–[:worksFor {role: 'engineer', from: 2000, until: 2010}]–>(b {name: 'BuildCo'}),
    (a)             –[:worksFor {role: 'manager', from: 2010}]–>(b) ✓

(3) CREATE (a {name: 'Adrian'})–[:worksAs {company: 'BuildCo'}]–>({role: 'engineer', from: 2000, until: 2010}),
    (a)             –[:worksAs {company: 'BuildCo'}]–>({role: 'manager', from: 2010}) ✓

(4) CREATE ({name: 'Adrian'})–[:worksAs {company: 'BuildCo'}]–>({role: 'engineer', from: 2000, until: 2010, role: 'manager', from: 2010}) ✗

**RDF* models**

(1)  <<<<:Adrian :worksFor :BuildCo>> :from 2000>> :role 'engineer' .
     <<<<:Adrian :worksFor :BuildCo>> :until 2010>> :role 'engineer' .
     <<<<:Adrian :worksFor :BuildCo>> :from 2010>> :role 'manager' .  ✓

(2)  <<<<:Adrian :worksFor :BuildCo>> :role 'engineer'>> :from 2000 .
     <<<<:Adrian :worksFor :BuildCo>> :role 'engineer'>> :until 2010 .
     <<<<:Adrian :worksFor :BuildCo>> :role 'manager'>> :from 2010 .  ✓

(3)  <<<<:Adrian :worksAs 'engineer'>> :for :BuildCo>> :from 2000 .
     <<<<:Adrian :worksAs 'engineer'>> :for :BuildCo>> :until 2010 .
     <<<<:Adrian :worksAs 'manager'>> :for :BuildCo>> :from 2010 .  ✓

(4)  <<<<:Adrian :worksAs 'engineer'>> :from 2000>> :for :BuildCo .
     <<<<:Adrian :worksAs 'engineer'>> :until 2010>> :for :BuildCo .
     <<<<:Adrian :worksAs 'manager'>> :from 2010>> :for :BuildCo .  ✓

**Figure 13.1 Modelling question illustrating modelling of class hierarchies**

### 13.1.2 Results and discussion

Figure 13.2 shows an analysis of the participants' responses. For both languages, a relatively high proportion of participants accurately identified the models as correct or incorrect. As before, some of the models may have been marked as incorrect because they were regarded as bad models, rather than being technically incorrect. Indeed, one participant commented that RDF* model 2 had been marked as incorrect because of its "flaws". Similarly, a participant commented of Cypher model 3: "Probably functionally correct given the description at the top, but you wouldn't want to bake the from/until values into the node. So "incorrect" because it's not scalable...". Indeed, looking at Figure 13.2, it is noticeable that, for both languages, when the questions are ordered by accuracy and by ranking, we arrive at the same ordering; this reinforces the view that participants were taking quality of model into account when assessing correctness.

Looking specifically at model rankings, Cypher models 1 and 4 are ranked very low, reflecting the fact that these models are incapable of answering the query. A few participants made comments about property values being overwritten. Of the two technically correct Cypher models, there is a clear preference for model 2, i.e. in which the role and timing metadata are included as edge properties. In the previous paragraph we noted a comment made regarding model 3, indicating a clear dislike for creating a node for the role and timing information. Related to this, one participant commented ":worksFor more intuitive (and useful) than :worksAs".

The intuition from the figure is confirmed by a statistical analysis. An ANOVA of rankings for the Cypher models showed a significant difference across the models ($F(3, 68) = 66.8$, $p < 0.001$). Moreover, a subsequent Tukey HSD analysis shows a significant difference in ranking between all the models, except models 1 and 4, i.e. the two incorrect models[29]. The significant difference between the two correct models indicates a firm preference for placing the metadata as properties in the predicate.

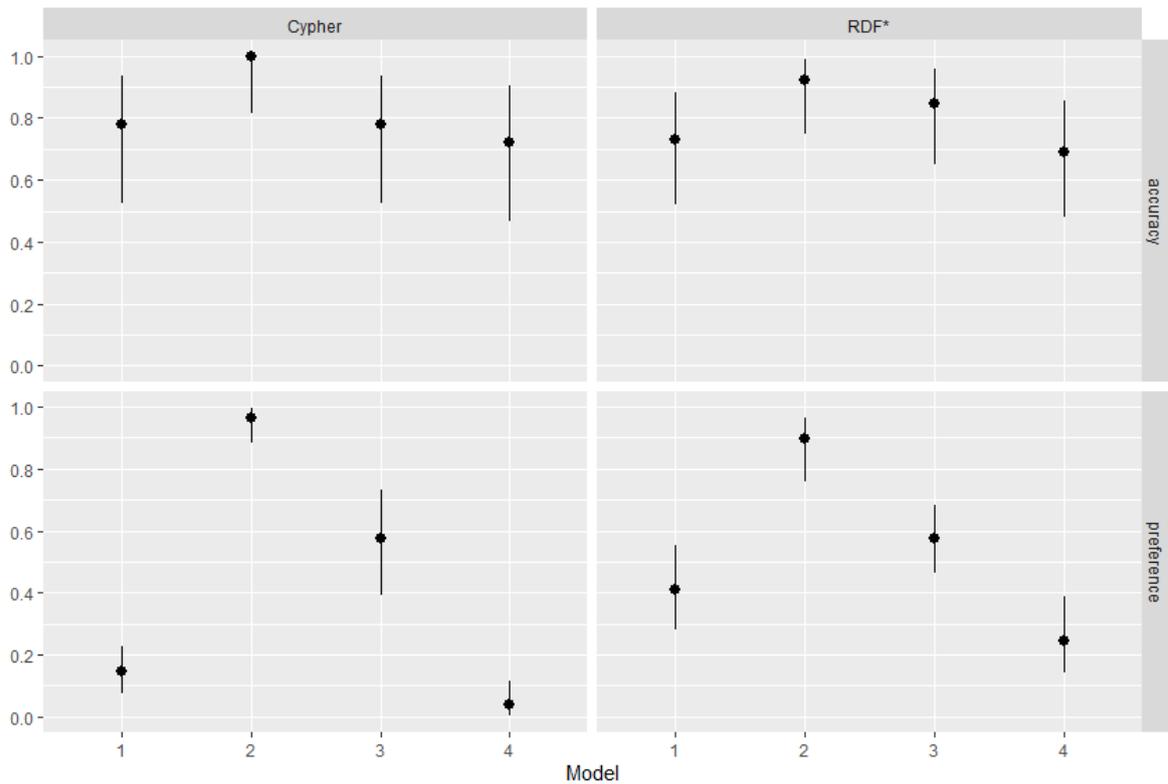

**Figure 13.2 Analysis of responses to modelling question shown in Figure 13.1**

There is also a considerable difference in ranking between the RDF* models, despite the fact that all these models are technically correct. The two most preferred models, models 2 and 3, both had the timing information in the outer triple. This may reflect that the two sentences in the English model had the timing information at the end. On the other hand, if the order of the English sentences was a dominant influence, we would have expected model 3 to be preferred to model 2. Possibly, the timing information seems more secondary, the company and the role being more fundamental; this is similar to the effect found in subsection 12.1.2. One participant commented on the difference between role and timing information:

> "There's a big difference between the properties "role" and "until": "role" qualifies the statement as additive information ("Adrian works for BuildCo, that's a fact, and *additionally* he is an engineer"), while "until" is subtractive: ("Adrian does *not* work for BuildCo since 2010"). There is no way to formally capture these differences in RDF*. One has to rely on human readable documentation to understand the differences in meaning."

It seems that human interpretation of the semantics is necessary to understand the model. For example, one participant commented, of model 2, but noting the problem was shared with other models, that if a person had the same role for the same company, "during two different periods of time", then it would not be possible automatically to correctly associate the start and end dates; this would require domain knowledge, i.e. about the properties of time. This problem does not arise with Cypher, since we can use two edges to represent the two separate periods of employment.

---

[29] $p < 0.001$ in all cases, except M1/M4: $p = 0.434$.

Another issue regarding the need for human interpretation is illustrated by the comment of one participant, who noted that model 1 "might give the impression that he stopped working at the company, since :until is specified for the :worksFor property".

Related to this point, the following triples, as in model 2, do not appear to be contradictory, since an individual can have two roles, either at the same time or different times:
    <<<<:Adrian :worksFor :BuildCo>> :role 'engineer'>> .
    <<<<:Adrian :worksFor :BuildCo>> :role 'manager'>> .

However, from model 1 we can extract two triples which, under human interpretation, are superficially inconsistent:
    <<<<:Adrian :worksFor :BuildCo>> :from 2000>> .
    <<<<:Adrian :worksFor :BuildCo>> :from 2010>> .

Effectively, to interpret doubly-nested triples, we need a clear view of the schema, i.e. the order of the metadata. Without that schema, the inner triples may, as in model 2, or may not, as in model 1, make sense on their own. We also need, when retrieving a triple, to retrieve all the associated metadata.

For the RDF* models, the intuition of Figure 13.2 is also confirmed by a statistical analysis. An ANOVA of the rankings shows a significant difference between the models ($F(3,100) = 21.59, p < 0.001$). A subsequent Tukey HSD indicated significant pairwise differences between all models, except models 1 and 3 and models 1 and 4[30]. In particular, model 2 was rated significantly better than each of the other three models.

Another perspective can be obtained by a two-factor ANOVA, where the factors are the position of the timing information, i.e. middle or outer triple; and the choice of predicate, or equivalently the object of the inner triple, i.e. company or role. There was a significant dependence on the position of the timing information ($F(1,100) = 46.673, p < 0.001$) and the choice of company or role within the inner triple ($F(1,100) = 16.454, p < 0.001$), with no significant interaction ($F(1,100) = 1.641, p = 0.203$). It appears that there are two non-interacting preferences: for timing information in the outer triple; and for company information, rather than profession, in the inner triple.

## 13.2 Querying

This question used model 2 from subsection 13.1, which was the preferred model for both sets of participants.

---

[30] M1/M2: $p < 0.001$; M1/M3: $p = 0.209$; M1/M4: $p = 0.209$; M2/M3: $p = 0.002$; M2/M4: $p < 0.001$; M3/M4: $p < 0.001$.

### 13.2.1 Question

Figure 13.3 shows the querying question. In both cases, the third query is correct; the other two are incorrect. For this question it proved difficult to create equivalent distractors in the two languages. The Cypher queries 1 and 2 wrongly assume the role and timing information is associated with the subject and object nodes respectively. The RDF* query 1 uses a join between two triples, rather than doubly-nesting the triples. The RDF* query 2 interchanges the position of the role and timing information.

**English query**

What were Adrian's roles at BuildCo, and when did he start them?

**Cypher model**

```
CREATE  (a {name: 'Adrian'})–[:worksFor {role: 'engineer', from: 2000, until: 2010}]–> (b {name: 'BuildCo'}),
        (a)                –[:worksFor {role: 'manager', from: 2010}]–>                (b)
```

**Cypher queries**

(1) MATCH (m {name: 'Adrian'})–[:worksFor]–>({name:'BuildCo'})   RETURN m.role, m.from   ✗

(2) MATCH ({name: 'Adrian'})–[:worksFor]–>(n {name:'BuildCo'})   RETURN n.role, n.from   ✗

(3) MATCH ({name: 'Adrian'})–[e:worksFor]–>({name:'BuildCo'})    RETURN e.role, e.from   ✓

**RDF* model**

```
<<<<:Adrian :worksFor :BuildCo>> :role 'engineer'>> :from 2000 .
<<<<:Adrian :worksFor :BuildCo>> :role 'engineer'>> :until 2010 .
<<<<:Adrian :worksFor :BuildCo>> :role 'manager'>> :from 2010 .
```

**SPARQL* queries**

(1) SELECT ?role ?date
    WHERE { <<:Adrian :worksFor :BuildCo>> :role ?role . <<:Adrian :worksFor :BuildCo>> :from ?date }   ✗

(2) SELECT ?role ?date
    WHERE { <<<<:Adrian :worksFor :BuildCo>> :from ?date>> :role ?role }   ✗

(3) SELECT ?role ?date
    WHERE { <<<<:Adrian :worksFor :BuildCo>> :role ?role>> :from ?date }   ✓

**Figure 13.3 Querying question using model 2 from subsection 13.1**

### 13.2.2 Results and discussion

Figure 13.4 shows an analysis of the participants' responses. It appears from the figure that the SPARQL* question were answered more accurately than the Cypher question. However, the lack of equivalence between the distractors, pointed out in the previous subsection, makes an overall comparison between the questions difficult. A safer approach is to compare the accuracy of responses to the two correct queries, query 3. A logistic analysis of deviance indicates a significant difference between the accuracy of responses for the two languages ($\chi^2(1) = 5.6839$, $p = 0.017$). However, a type II two-factor analysis of deviance suggests that, after controlling for the participants' knowledge of the respective languages, there is no significant difference between the two languages (language: $\chi^2(1) = 1.2953$, $p = 0.255$), nor between the levels of previous knowledge ($\chi^2(3) = 2.4377$; $p = 0.487$). There was also no significant interaction effect ($\chi^2(2) = 0.0000$, $p = 1.000$).

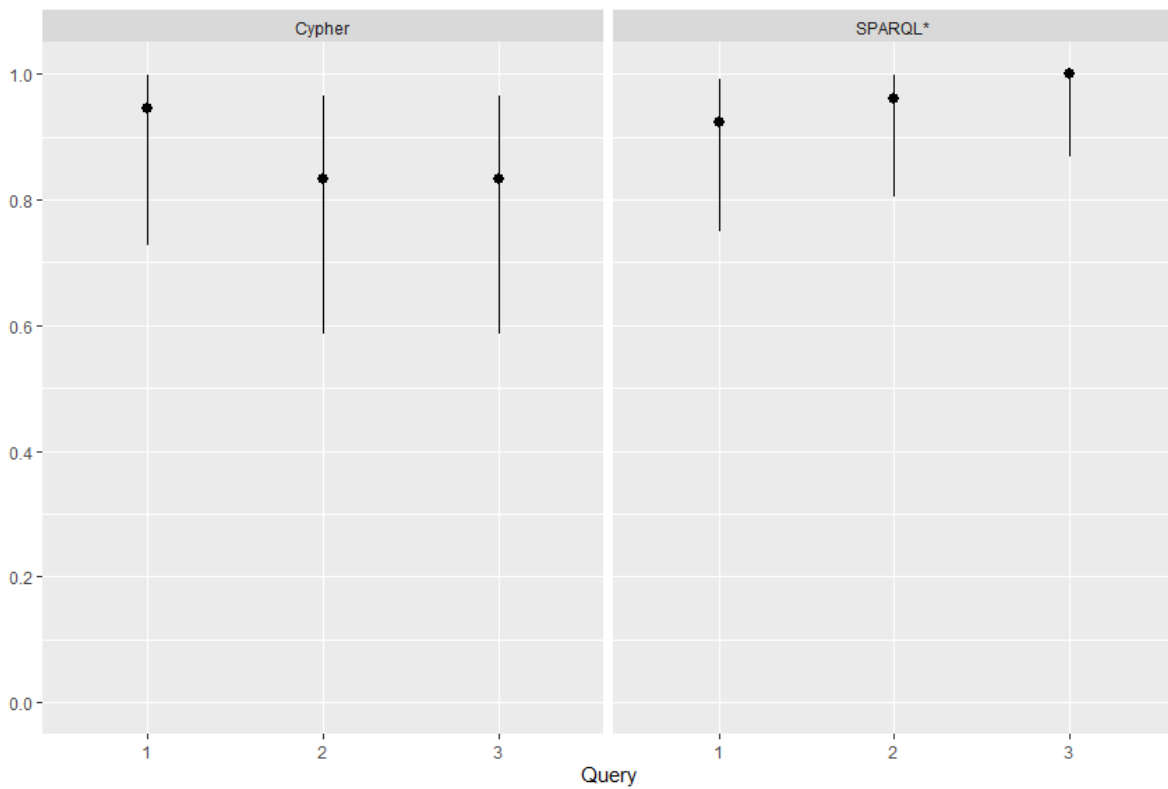

**Figure 13.4 Analysis of responses to modelling question shown in Figure 13.2**

# 14 Concluding discussion

In this section, we summarise the main conclusions from our study and briefly review current work to extend graph database languages.

## 14.1 Summary of study findings

At the beginning of the paper we set out three objectives:
- To compare the ease of use of the two language paradigms.
- To identify participants' preference for alternative semantically equivalent models; and, where appropriate, whether those preferences differed between the two languages.
- To determine where participants had particular difficulties in interpreting and querying models.

### 14.1.1 Comparison of the two paradigms

With one exception, described in subsection 9.2, there was no evidence that, for the examples given, one of the language paradigms was significantly easier to use than the other. This is consistent with earlier studies of programming languages. For example, Green et al. [28] suggest that user training may be more important than the particular notation; and that different notations may better suit different tasks. What we do see in certain situations, as in Sections 10 and 11, is a preference for a similar modelling style in both languages.

However, our results need to be interpreted carefully. There were limitations in our methodology, discussed in subsection 15.2, which may have reduced our ability to detect such differences.

### 14.1.2 Alternative modelling approaches

Where the Cypher and RDF* models were analogous, there was, with one exception, agreement as to the order of preference. This was in subsection 12.1, which we discuss below.

In subsection 9.1, where there was only one correct Cypher model, this was by far the preferred model. For RDF*, where all models were correct, there was a preference for the model analogous to the correct Cypher model. This was the RDF* model in which the role information was the object of the outer triple, rather than being associated with the person or the company.

In subsection 10.1, there was a preference for the model which, in Cypher used a node for each city, and in RDF used an IRI. However, this preference was significantly greater in RDF than in Cypher.

In subsection 11.1, where we considered class hierarchies, we saw that for both languages participants preferred the model which did not require manipulation of text; although this preference was significantly less for the Cypher participants. Consistent with this preference, participants were better able to identify correct and incorrect queries in subsection 11.3, where textual manipulation was not required, than in subsection 11.2, where it was necessary.

Subsection 12.1, which described models with metadata about node metadata, illustrated the one exception to our comment at the start of this subsection. For RDF* there was a clear preference for models in which the date, rather than population, was in the outer triple; corresponding, perhaps, to our natural way of thinking about the order of the metadata. For Cypher there was no significant difference in preference between a model in which the date information was in the edge and the population information in the node, and vice versa; suggesting that both seem equally natural within the property graph paradigm. One participant did suggest the need for composite properties, so that property / value pairs could be identified as being associated; another suggested the use of a node representing population size and date.

In subsection 13.1 we again saw, for the RDF* model, a preference for the date to be in the outer triple, on this occasion rather than role or company information. For the two correct Cypher models, we saw a clear preference for putting the role and date as edge properties, whilst the company constituted a node. This is the obvious modelling strategy, opening up the possibility of adding further information about the company.

In general, the preference for models confirmed our intuitions. This is borne out by the fact that, where a modelling question was followed by one or more querying questions, then in the latter we had used the model to which participants gave the highest preference. The only exception to this was in Section 11, where we used both models in the subsequent querying questions.

### 14.1.3 Particular difficulties with models

We have noted that the accuracy of understanding of the models was difficult to gauge, because in answering the questions regarding correctness or incorrectness of models, some participants were influenced by their assessment of the quality of the models. For the queries, with the exception of those discussed in subsection 11.2, there was a high level of accuracy in identifying correct and incorrect queries. The exception was where there was a need to perform string manipulations using built-in functions in Cypher and SPARQL. Here, participants had considerable difficulty. This was true for both sets of participants; after controlling for previous language knowledge, there was no significant difference between the two languages.

## 14.2 Evolving graph database languages

SPARQL and the various query languages for property graphs occupy different worlds. The former has been standardised by a public process, organised by the W3C. The latter have been less formally developed, e.g. Neo4j's Cypher. Nevertheless, in both cases there is a public debate about the development of the languages. There is currently a W3C SPARQL1.2 Community Group[31] which is collecting issues[32]. These issues range widely, e.g. consideration of SPARQL*; incorporation of units, such as for temperature; additional aggregate functions. Similarly, there is a process for proposing improvements to Cypher[33].

Rather than consider the range of proposals being put forward, we concentrate here on two fundamental issues which relate to aspects of our study: reification and query-time reasoning.

### 14.2.1 Reification

RDF* and SPARQL* were introduced by Hartig and Thompson [5] to provide an improved approach to reification. However, their approach is not without difficulties. We saw in Section 13, that when using multiple nesting of RDF* triples, one needs to be aware of the order of nesting, e.g. whether the role or time is in the outer triple. This problem can be avoided in Cypher, since the role and time can be together associated with an edge or a node. Moreover, Cypher permits several distinct edges with the same label between the same two nodes. Although it is probably natural to think of time as qualifying role, in the Cypher model this need not be explicit; one can think of time and role simply as being associated. In any case, in Cypher we do not need to be aware of an order, but only the property names.

The Stardog[34] knowledge graph has implemented RDF* and SPARQL*, but in a way which also permits edge properties with a similar syntax to Cypher. An example of RDF* given in the Stardog documentation is:

    << :Pete a :Engineer >>           :since  2010 .
    << :Pete :worksAt :Stardog >>    :source :HR .

However, this can also equivalently be written using edge properties:

---

[31] https://www.w3.org/community/sparql-12/
[32] For the list of issues, see: https://github.com/w3c/sparql-12/issues/
[33] https://groups.google.com/g/opencypher/c/SCSRlplb_zc
[34] https://www.stardog.com/    For more detail on the approaches described here, see: https://docs.stardog.com/query-stardog/edge-properties

```
    :Pete   a { :since 2010 }           :Engineer ;
          :worksAt { :source :HR }      :Stardog
```

The latter format remains RDF, but adopting the edge property syntax of property graphs. Note that this example includes both kinds of metadata discussed in subsection 3.2. The *:source* metadata is metadata about the triple; the *:since* metadata adds information to the triple.

The implementation of RDF* is subject to a number of restrictions. Two of these are amongst the restrictions listed in subsection 4.1: triples can only be embedded as the subject of a metadata triple, i.e. not predicate or object; and metadata triples cannot be nested. These restrictions help make possible conversion to the edge property syntax.

A query to determine who works at Stardog as an engineer, since when and the provenance of the employment information, could be written using a SPARQL* syntax:
```
    SELECT * {
        << ?emp a :Engineer >>           :since ?year .
        << ?emp :worksAt :Stardog >>     :source ?who .
    }
```

However, the SPARQL query could also be written using edge properties:
```
    SELECT ?emp ?start ?end ?who {
        ?emp a { :since ?start }         :Engineer ;
            :worksAt { :source ?who }    :Stardog .
    }
```

It is also possible to have multiple edge properties in Stardog:
```
    :Pete   a { :since 2010 ; until 2018}  :Engineer ;
          :worksAt { :source :HR }       :Stardog
```

Similarly, multiple edge properties can also be used in SPARQL queries.

Another difficulty with RDF* reification concerns the implementation practice of asserting embedded triples. Commenting on the RDF* model 1 in subsection 9.1, one participant commented that the query could only be answered by "ignoring other kinds of metainformation, such as, for instance, that 'Adrian worksFor X' is a false claim by Bob". In fact, an RDF* model needs to avoid any sort of negation of embedded triples. This issue extends beyond explicit negation. Confidence levels are often quoted as an example of triple metadata. However, low confidence levels are close to denials of the triple. Similarly, where the provenance of a triple is a source not regarded as highly trustworthy, this also casts doubt on the validity of the triple. The underlying problem is the possibility of using a query to retrieve embedded triples without the associated metadata; thus one may retrieve a triple without, e.g., retrieving its confidence level.

A more radical approach to property graph databases has been developed by Grakn Labs[35]. They use the hypergraph data model [29]. Whereas an edge in an ordinary graph connects two nodes, a hyperedge in a hypergraph can connect any number of nodes. Thus, n-ary relationships are a native feature, rather than required to be built up from binary relationships. Hyperedges can also connect other hyperedges. This can be regarded as an extension of the RDF concept that a predicate, since it is represented by an IRI, can be regarded as the subject or object of another predicate. For a case study of the use of a Grakn knowledge graph in an engineering application, see Berquand and Riccardi [30].

---

[35] https://blog.grakn.ai/

### 14.2.2 Query-time reasoning

We are concerned here with the simplest type of reasoning, based on hierarchies of classes and predicates. RDFS provides two predicates, *rdfs:subClassOf* and *rdfs:subPropertyOf*, which enable class and predicate hierarchies to be constructed. These can be used in two ways. Inferencing can be used to achieve logical closure at load-time. However, even for closed systems load-time closure is not always possible, or at least efficient. Moreover, it is not possible for linked open data. There is a need for hierarchical reasoning at query-time. Section 11 provided a use-case for class hierarchies, and showed how query-time reasoning can be achieved with SPARQL. A similar approach can be taken for query-time reasoning with predicate hierarchies, using *rdfs:subPropertyOf*. This approach is currently not possible where a predicate forms part of a predicate path, since variables are not permitted in these paths. However, the inclusion of variables in predicate paths has been proposed as an extension to SPARQL[36]. Peréz et al. [31], in a paper which proposed a navigational language for querying RDF, prior to the introduction of SPARQL predicate paths, provide use-cases for both class and predicate hierarchies, including where those predicates form part of predicate paths.

Cypher has no built-in equivalents of *rdfs:subClassOf* and *rdfs:subPropertyOf*. However, Section 11 also showed how class hierarchies could be taken account of in Cypher querying, either by emulating *rdfs:subClassOf* or by using an approach based on Cypher labels. For edge hierarchies, we cannot emulate the approach of *rdfs:subPropertyOf* because this depends on the use of IRIs to represent predicates, which means that RDF predicates can also be regarded as nodes. In Cypher, edges and nodes are distinct. This rules out an approach analogous to that of model 2 of subsection 11.1, as illustrated by Cypher query 3 in subsection 11.3. However, it would be possible to adopt an approach analogous to that of model 1 in subsection 11.1, as illustrated by Cypher query 1 in subsection 11.2. In place of the *label()* function used in that query, one could use the *type()* function to return the string representation of an edge type. However, as we have seen in Section 11, this approach is cumbersome, was not liked by participants, and led to difficulties in identifying the correct query.

An alternative approach would be to extend Cypher to permit the definition of node and type hierarchies, i.e. with predefined relationships equivalent to *rdfs:subClassOf* and *rdfs:subPropertyOf*. These hierarchies could then be invoked automatically at query-time. This is, in fact, the approach taken by Grakn Labs. They permit the creation of hierarchies for entities (i.e. the equivalent of Cypher nodes), relations (Cypher relationships) and attributes (Cypher properties). Query-time reasoning takes account of these hierarchies. Unlike with RDFS, any entity, relation or attribute can have only one parent. The system also permits the creation of rules, which enables more complex query-time reasoning. The case study mentioned in the previous sub-section, by Berquand and Riccardi [30], includes the use of query-time reasoning with rules.

---

[36] https://github.com/w3c/sparql-12/issues/65

# 15 Recommendations

In this section we make three sets of recommendations: about modelling, about research methodology, and about future research directions. In subsection 15.1 we draw on our results, and on comments made by study participants, to make some recommendations about modelling practices. In subsection 15.2 we recommend how future work with an extended methodology could build on our initial study; to overcome the limitations which may have reduced our ability to detect differences between the language paradigms. In subsection 15.3 we take account of the development of graph database languages to recommend future areas for investigation.

## 15.1 Modelling

Firstly, we offer some recommendations which are relevant to both modelling paradigms:
- In the choice between creating a new node or using a literal, use a node where this is likely to support future expansion of the model; use a literal where the resource it represents is not likely to enter into relationships with other resources or to have properties. Thus, in general, use a node to represent a city, a literal to represent a number. On the other hand, if our knowledgebase is about numbers, and the properties of numbers, we might want to represent numbers as nodes.
- Avoid modelling decisions which necessitate queries involving conversion from text string to resource name, as was necessary in subsection 11.2.
- Carefully choose predicate names so as to obviously link their subjects and objects in a sensible way.

In the property graph paradigm:
- Where properties describe the relationships between nodes, they should be associated with the edge joining the nodes, rather than the nodes themselves. For example, where we are concerned with an individual's employment in a company, role and date properties are best associated with the edge between the individual and the company.
- Where possible, consider the use of composite properties, e.g. where a date needs to be associated with a population size. Where this facility is not available, consider the use of nodes with multiple properties.

When using RDF*:
- Think carefully about the consequences of the assertion of inner triples; in particular where inner triples are negated, or partially negated by the use of a low confidence rating or having a dubious provenance.
- Where you have several levels of nesting, think carefully about the order of nesting. Where metadata appears to qualify other metadata, then it is more natural for the former to be at the outer level of nesting. For example, where we are talking about a person's role at a given date, it is more natural for the date information to be at the outer level.
- As a further comment on the previous point, avoid orders of nesting which appear to create inconsistencies. Inner triples which together ascribe multiple roles to an individual appear less inconsistent than inner triples which ascribe multiple dates to the same event, e.g. commencement of employment. See the discussion in subsection 13.1.2

## 15.2 Extending the methodology

There are three ways in which our research methodology could be extended to better differentiate between the usability of the two paradigms.

Firstly, a future study could use more complex models and queries. Many of our querying questions were answered with a high degree of accuracy. More demanding questions might differentiate between the languages better.

Secondly, differentiation between the languages would be aided by recording how long participants took to answer each question. In studies of this kind, it is possible that where accuracy does not differentiate between approaches, response time will. Our approach was designed to reach out to a wide community of people with some experience of graph database languages; rather than working, e.g. in our own institution, with people who might not have had any such expertise. The downside of our remote, automated approach was that we were not able to record response times. A future study could use a supervised approach to remedy this. The use of response time data might particularly help to differentiate between the iconic approach of Cypher, using forward and backward arrows, with the symbolic approach of SPARQL, using ^ to represent reverse directionality in predicate paths.

Thirdly, a comparison between the languages would be aided by a better balance between the sets of participants. In our study, the RDF* / SPARQL* participants were considerably more experienced than the Cypher participants. All the RDF* / SPARQL* participants had at least a little knowledge of SPARQL, and 69% had expert knowledge; whilst 39% of the Cypher participants had no knowledge of Cypher. Ideally, future studies should be focussed on experienced users. Where this is not possible, at least a similar distribution of user experience should be aimed for.

## 15.3 Future research directions

In comparing RDF* and SPARQL* with Cypher, our study has focussed particularly on reification. We have also investigated approaches to query-time reasoning. These are both areas for further research.

In this paper we have discussed three approaches which either claim to offer improved RDF reification or offer alternative modelling approaches:
1. The use of RDF* and SPARQL*, as discussed in this paper. These extensions to RDF and SPARQL were proposed by Hartig and Thompson [5] as an alternative to the previous cumbersome techniques for reification, e.g. as identified by Hernández et al. [11].
2. The use of edge properties in property graphs, which provide metadata about relations. Included here is the hybrid Stardog approach, where edge properties can be included in RDF and SPARQL, in a such a way that translation is possible into RDF* and SPARQL*.
3. The hypergraph data model, as implemented by Grakn, in which reification is not necessary since relationships are natively n-ary.

These approaches need to be compared from the standpoint of usability. This study has concentrated on comparing (1) and (2). We have already commented on the need to extend the study by using realistically complex models, and response time data to better differentiate approaches. Our comparison should also be extended to (3). An important question is to what extent the different approaches offer advantages in different modelling situations; and to what extent there is a trade-off between the advantages offered and increased complexity of approach.

Query-time reasoning was explored in Section 11 and discussed further in subsection 14.2.2. For RDF models it is made possible by the use of *rdfs:SubClassOf* and *rdfs:SubPropertyOf*. We have seen that there is still a limitation caused by the inability to include a variable in a SPARQL predicate path. For the property path model, e.g. as implemented in Cypher, the limitations are greater, caused by the inability to define hierarchies of node labels and edge types. The hypergraph model, as implemented by Grakn, offers the ability to define hierarchies of entities, relation and attributes. In addition, the hypergraph model offers the ability to create relations between relations, since hyperedges can be treated as nodes and linked by hyperedges. As with reification, we need to compare different approaches on real use-cases.

## Acknowledgements

The authors would like to thank all those participants who took part in the study.